\documentclass[11pt, draftcls, peerreview]{IEEEtran}
\usepackage{cite,amsmath,amssymb,graphicx}
\usepackage{citesort}
\usepackage{algorithm,algorithmic}
\usepackage{subfigure}
\usepackage{mathrsfs}
\usepackage{color}
\usepackage{url}
\usepackage{bbm}
\usepackage[T1]{fontenc}% to avoid type 3 font: dvips -t letter -Ppdf -G0 -j0 %mypaper.dvi -o mypaper.ps

\newcommand{\E}{\mathbb{E}}
\newcommand{\Y}{\mathbf{y}}

\newcommand{\R}{\mathbf{r}}

\newcommand{\uD}{\overline{\Delta}}
\newcommand{\lD}{\underline{\Delta}}
\newcommand{\res}{\varepsilon}
\newcommand{\ignore}[1]{{}}
\begin{document}
\title{Cooperative Change Detection for
Online Power Quality Monitoring}
\author{Shang Li$^\dagger$ and Xiaodong Wang$^\dagger$
\thanks{$^\dagger$S. Li and X. Wang are with Electrical Engineering Department, Columbia University, New York, NY 10027 (e-mail: \{shang,wangx\}@ee.columbia.edu).
}}
\maketitle
\begin{abstract}
This paper considers the real-time power quality monitoring in power grid systems. The goal is to detect the occurrence of disturbances in the nominal sinusoidal voltage/current signal as quickly as possible such that protection measures can be taken in time. Based on an autoregressive (AR) model for the disturbance, we propose a generalized local likelihood ratio (GLLR) detector which processes meter readings sequentially and alarms as soon as the test statistic exceeds a prescribed threshold. The proposed detector not only reacts to a wide range of disturbances, but also achieves lower detection delay compared to the conventional block processing method. Then we further propose to deploy multiple meters to monitor the power signal cooperatively. The distributed meters communicate wirelessly to a central meter, where the data fusion and detection are performed. In light of the limited bandwidth  of wireless channels, we develop a level-triggered sampling scheme, where each meter transmits only one-bit each time asynchronously. The proposed multi-meter scheme features substantially low communication overhead, while its performance is close to that of the ideal case where distributed meter readings are perfectly available at the central meter.
\end{abstract}

\begin{IEEEkeywords}
Power signal disturbance, autoregressive model, change detection, level-triggered sampling.
\end{IEEEkeywords}

\section{Introduction}

At present the power quality has become a critical security concern for the emerging power grid system, due to the rapidly growing number of equipments that not only generate but also are sensitive to various disturbances.
One crucial task is monitoring the power signal for malicious power quality disturbances that could lead to device damage or even network blackout.
In general, power quality disturbances include the voltage disturbance and the current disturbance. They both involve deviations of the actual power signal from the nominal sinusoidal waveform with prescribed amplitude and frequency, thus can be treated similarly from the signal processing point of view. In this paper, without loss of generality, we consider the voltage disturbance. The real-time disturbance detection is useful in practice in two ways. On one hand, it enables the power system to promptly respond to the detrimental fluctuations caused by the generator and load operations, capacitor bank switching and abrupt environment change (e.g., lightning strike). On the other hand, it also serves as an abnormal data recording trigger. In power grids, it becomes more and more important to record the power quality data for off-line assessment, in order to help the electricity providers to improve their power supply. However, to avoid the huge volume of data storage, the recorder should only capture the informative data segments corresponding to abnormalities or disturbances and not record data when the power signal is normal. Therefore, to trigger the recording process, timely detection of the disturbance is required.

The monitoring procedure is commonly realized by sampling and analyzing in real-time the  voltage waveform, which in practice, is always corrupted by noise. If the  noise level is comparable to that of the disturbance, it could lead to frequent false alarms or large decision delay, resulting in poor detection performance. To that end, one key task in power quality monitoring is to detect the voltage disturbance from the observed noisy waveform as soon as possible after its occurrence, so that certain protection measures can be taken and/or the recoding process can be triggered immediately.

\subsection{Background}
The existing disturbance detection methods  can be roughly categorized into non-model-based and model-based approaches \cite[Chapter 7]{Bollen_book} \cite{Bollen09}. In particular, the non-model-based approach directly examines the instantaneous changes from the observed  waveform. For example, one method is to use a high-pass filter to capture the high-frequency component induced by the abrupt transition when the disturbance occurs. Intuitively, this method becomes ineffective if the waveform changes smoothly. Moreover, the high-frequency component at the transition point can be buried by the noise. The most widely used method thus far is by monitoring the root mean squared (RMS) sequence, which is computed over a sliding window of length $W$ (usually one cycle of the nominal waveform) as follows:
\begin{align}\label{def:RMS}
  Q(t)=\sqrt{\frac{1}{W}\sum_{k=t-W+1}^t y_k^2}\;,
\end{align}
where $y_k$ is the $k$th sample of the voltage waveform. A disturbance is detected once the current RMS surpasses or falls below a prescribed threshold. Despite its simplicity, the RMS method is effective in detecting the disturbances associated with magnitude change, e.g., voltage sag, voltage swell. However, it is shown in \cite{Garoom05} that the RMS method can miss some transient disturbances that are associated with spectral changes, which is expected because it is based on the energy level of the waveform and is less sensitive to spectral variation.
Other methods detect the instantaneous distortion in the frequency domain, mainly by the wavelet transform (WT) or the short-time Fourier transform (STFT) \cite{Dwivedi09,Bollen09}. These methods are naturally more sensitive to the spectral distortion. In addition, both RMS and STFT (or WT) methods have limited time resolution due to the window over which the RMS and spectrum are evaluated respectively.

In contrast, the model-based framework detects the change of parameters or the large residual between the observed waveform and the nominal model. A typical model is to characterize the nominal waveform and disturbance by the superposition of a number of sinusoidal waveforms.
Then classical spectral estimation methods such as the MUSIC, ESPRIT and Kalman filter can be applied to estimate the parameters of these sinusoidal disturbances \cite{Dafis00,Bollen_book}.
The first two methods both require an observation window to estimate the parameters, thus decreasing the time resolution of the detection. All three methods heavily rest on the presumed model, e.g., they require the knowledge of the number of sinusoidal waveforms in the model, which makes them less robust. Alternatively, the autoregressive (AR) model is employed for the voltage disturbance detection in \cite{Gu00}, which is capable of capturing a broad range of spectral property, whereas the sinusoidal model can only capture a certain fixed number of frequency components. However, the method in \cite{Gu00} is based on examining the residual during the waveform transition, making it vulnerable to noise.

An effective approach to mitigating the noise is to employ the statistical framework of hypothesis testing. A block-sequential Neyman-Pearson test is used in \cite{Gu04} to analyze the cause of the voltage disturbance \cite{Gu04}. That work treats the off-line disturbance classification rather than the online disturbance detection. In \cite{Dwivedi09}, the Kolmogorov-Smirnov (KS) test and the likelihood ratio test are employed to de-noise the wavelet transform coefficients. It is a fixed-sample size approach rather than a sequential one, thus is less efficient in terms of time resolution. In \cite{Xingze10,Xingze12}, under the change detection framework, a sequential online approach based on the weighted CUSUM test was introduced, by examining the different distributions of the observed waveforms before and after the occurrence of the disturbance. However, the disturbance signal is treated as independent over time.

Another noteworthy approach to combat the noise is to employ cooperative meters. In reality, disturbances tend to occur to a group of connected electrical buses at the same time, which brings about the opportunity of detecting the disturbance occurrence in a collaborative fashion. That is, by employing multiple meters at these connected electrical buses, one can draw on the diversity across meters to achieve better detection performance than using a single meter \cite{LiCAMSAP13}. In the context of deploying wireless Cyber-network for power system monitoring, meter installation and communication protocols are discussed in \cite{Hoglund12,Ilic12}. However, specific signal processing techniques of cooperative disturbance detection on top of these physical infrastructures is yet to be investigated.

\subsection{Overview}
In this paper, we also formulate the online voltage monitoring as a sequential change detection problem. But compared to \cite{Xingze10}, we build our framework  on the time series model, i.e., AR model, which captures the time correlation of the disturbance, and thus provides more realistic characterization.
To tackle the main difficulty that the disturbance signal is typically unknown, we propose a change detector based on the generalized local likelihood ratio (GLLR) test. This approach takes a simple form and is more effective than the existing methods.

Moreover, we consider the scenario where multiple meters are employed for cooperatively monitoring the voltage waveform. As mentioned before, these meters are deployed at various locations in the distribution network, where electrical buses are exposed to the same power quality event\footnote{Although the power quality event strikes a group of buses altogether, the disturbance signals observed at these buses are not necessarily the same. For example, a line fault event could result in voltage sag at one bus but voltage swell at another bus.}.
{Despite the robustness against noise, distributed meters inevitably impose communication challenges. Conventionally, meters transmit their local measurement (quantized with multiple bits) to the central meter, where the cooperative detection is performed. The ideal case is that the distributed meter readings are precisely available to the central meter at every sampling instant, i.e., infinite number of bits  for quantization, referred to as the centralized multi-meter detection.  However, such a centralized setup induces significant amount of communication overhead. To meet the bandwidth constraints, we develop a decentralized detection scheme where it is assumed that the distributed meters and the central meter are linked by low-rate communication channels.} In particular, in the proposed decentralized scheme, each meter performs its own GLLR test; samples its decision statistic using the level-triggered sampling; and transmits its sampled local statistic to the central meter using a single bit. The central meter collects the bits from all linked meters and updates the global decision statistic to make the decision.  In Section IV, we provide extensive simulation examples to demonstrate that the proposed decentralized detector performs close to the centralized detector, and outperforms the traditional decentralized approach that is based on uniform-in-time sampling and quantization.

The reminder of this paper is organized as follows. In Section II, we formulate the disturbance detection as a sequential change detection problem based on the AR model and develop a generalized local likelihood ratio (GLLR) test. In Section III, we further develop the decentralized GLLR test based on the level-triggered sampling. Simulation results are provided in Section IV and finally, Section V concludes the paper.

\section{Sequential Change Detection of Disturbances based on the AR Model}

In this section, we formulate the online detection of voltage disturbance as a sequential change detection problem with unknown post-change parameters. Then  a  generalized local likelihood ratio-based detector is derived.
\subsection{AR Modelling of Disturbances}
Without loss of generality, we assume that the disturbance occurs at some unknown time $t=t_0$. That is, before $t_0$ the nominal voltage waveform is a sinusoid with some nominal magnitude, frequency and phase:
\begin{align}\label{nominal}
  f^{\text{nominal}}_t=a_0\sin(2\pi f_0t+\phi_0).
\end{align}
After $t_0$, the disturbance distorts the nominal sinusoid. Since the parameters $\{a_0, f_0, \phi_0\}$ take prescribed values ($\phi_0$ is obtained by synchronization), the waveform before $t_0$ is deterministic. Thus we can subtract the nominal waveform (2) from the measurement to obtain a signal consisting only of noise and disturbance \cite{Xingze10,Xingze12}. This preprocessing procedure isolates the disturbance signal, which is comparatively weaker than the nominal waveform \cite{Shin06}.
Consequently, the post-precessing observed signal before the disturbance occurs consists of measurement noise, which is typically modeled as a white Gaussian process, i.e.,
\begin{align}\label{before_change}
y_t=\nu_t\sim {\cal N} (0, \sigma_\nu^2), \quad t<t_0.
\end{align}
where $y_t$ is the post-preprocessing meter observation at sampling instant $t$. After $t_0$,  the meter observations are comprised of disturbance signal and the measurement noise. To capture the time correlation of the disturbance, we propose to use an autoregressive (AR) model to characterize the disturbance signal, which is popular in analyzing the spectral property of various types of signals, such as speech signals \cite{Obrecht88} and seismic signals \cite{Tjostheim75}, and is applied to model power signal in \cite{Gu00}.
The AR model is able to represent a broad spectral range, yielding robust characterization of a variety of potential disturbances; whereas the sinusoidal  model  (i.e., modeling the disturbance signal as a sum of sinusoids) in \cite{Dafis00} only represents a fixed number of certain frequency components. Moreover, the AR model also requires a small number of parameters and its performance is robust to the model order $p$ \cite{Basseville}. In particular, the post-preprocessing signal after the disturbance occurs can be modeled as
\begin{align}\label{model}
\left\{
\begin{array}{l}
y_t=x_t+\nu_t, \\
x_t = \tilde{\mu}+\sum_{j=1}^p a_j\left(x_{t-j}-\tilde{\mu}\right)+w_{t}, \quad t\ge t_0,
\end{array}\right.
\end{align}
where $x_t$ is disturbance signal modeled by an AR process with mean $\tilde{\mu}$, $w_t$ is the driving noise of the AR process and $y_t$ is the meter observations. We further write \eqref{model} as
\begin{align}\label{change_model}
y_t=\mu+\sum_{j=1}^pa_jy_{t-j}+u_t, \quad  t\ge t_0,
\end{align}
where $\mu=\tilde{\mu}\left(1-\sum_{j=1}^pa_j\right)$, and $u_t\triangleq \left(\nu_t-\sum_{j=1}^pa_j\nu_{t-j}+w_t\right) \sim {\cal N} (0, \sigma_u^2)$ with $\sigma_u^2=(1+\sum_{j=1}^p({a_j^t})^2)\sigma_v^2+\sigma_w^2$ accounts for the excitation of the disturbance and the measurement noise.

Note that the statistical models before and after the disturbance, i.e., \eqref{before_change}-\eqref{change_model}, correspond to a standard change detection formulation. The change detection (also termed as the quickest detection \cite{Poor_book}) aims at detecting the change point as quickly as possible. It achieves high time-resolution by sequentially observing the measurements in the time domain, and the detection delay is minimized subject to a false alarm constraint. The optimal algorithm is obtained by finding the stopping time $T$\footnote{A stopping time is a random variable, whose value $\{T=t\}$ is determined by the random samples up to $t$.} such that
\begin{align}\label{change_detection}
\inf_{T}\;\sup_{t_0}\; \E_{t_0}\left(\left(T-t_0\right)^{+}|T\ge t_0\right)\quad
\text{subject to}\;\; \E_{\infty}\left(T\right)\ge \gamma,
\end{align}
where $\E_{t_0}$ means the expectation given the change point at $t_0$, and $\E_\infty$ the expectation without any change point. Therefore, the objective function in \eqref{change_detection} corresponds to the average detection delay and $\E_\infty(T)$ corresponds to the false alarm period, i.e., the time a false alarm appears. Intuitively, \eqref{change_detection} aims to minimize the mean detection delay while controlling the period before a false alarm to be longer than $\gamma$.

Note that in the change detection formulation, it is assumed that the post-change event is ever-lasting. Although in practice some power quality disturbance is  transient in time (i.e., \eqref{change_model} holds for  $t_0 < t < t_1$, and after $t_1$, the system resumes the normal condition \eqref{before_change}), the expected detection delay is supposed to be much less than the transient duration (such that timely protection measures can be taken before any damage is incurred). Thus we assume $t_1$ is sufficiently large such that the sequential change detection framework is valid. In Section V, we provide simulation examples to illustrate the performance of the proposed detectors on transient disturbances.

Denoting the parameter vector comprised of the AR coefficients and the variance as
${\boldsymbol\theta}=[a_1, a_2, \ldots, a_p,\mu ,\sigma]^\intercal$,
then the waveform change in \eqref{before_change} and \eqref{change_model} corresponds to the change of the parameter vector at $t=t_0$:
\begin{align}\label{def:hypothesis}
\left\{\begin{array}{ll}
{\boldsymbol\theta}={\boldsymbol\theta}_0\triangleq [0, 0, \ldots, 0, 0, \sigma_\nu]^\intercal, & \quad t<t_0, \\ {\boldsymbol\theta}={\boldsymbol\theta}_1\triangleq  [a_1, a_2, \ldots, a_p, \mu, \sigma_u]^\intercal, & \quad t \geq t_0.
\end{array}\right.
\end{align}
Further denoting ${\Y_j^k}\triangleq [y_k, y_{k-1}, \ldots, y_{j}]^\intercal$,
the joint conditional probability density function at time $t$ under these two parameter vectors can be expressed as
\begin{align}\label{LLR}
&f_{\boldsymbol\theta}(y_t|{\Y_1^{t-1}})\nonumber\\=&
\left\{
\begin{array}{c}
\frac{1}{\sqrt{2\pi}\sigma_\nu}\exp\left(-
\frac{{\varepsilon_{t,{\boldsymbol\theta}_0}}^2}{2\sigma_\nu^2}\right), \quad {\boldsymbol\theta}={\boldsymbol\theta}_0\\
\frac{1}{\left(\sqrt{2\pi}\sigma_u\right)}\exp\left(-
\frac{{\varepsilon_{t,{\boldsymbol\theta}_1}}^2}{2\sigma_u^2}\right), \quad {\boldsymbol\theta}={\boldsymbol\theta}_1
\end{array}\right.,
\end{align}
with $\varepsilon_{t,{\boldsymbol\theta}_0}\triangleq y_t$ and $\varepsilon_{t,{\boldsymbol\theta}_1}\triangleq y_t-\mu-\sum_{j=1}^pa_jy_{t-j}$.
Then the occurrence of the disturbance can be detected using the following sequential change detection procedure:
\begin{align}\label{general_CUSUM}
&g_k=\max_{1\leq j\leq k} S_{j}^k,\quad T=\inf \{k: g_k \geq h\},
\end{align}
where $h$ is a decision threshold and
\begin{align}\label{LLR_N}
S_{j}^k
&\triangleq \log\frac{f_{{\boldsymbol\theta}_1} ({\Y _{j}^{k}}|{\Y_1^{{j}-1}})}{f_{{\boldsymbol\theta}_0} ({\Y _{j}^{k}}|{\Y_1^{{j}-1}})}=\sum_{i={j}}^k \log \frac{f_{{\boldsymbol\theta}_1} (y_i| {\Y _{i-p}^{i-1}})}{f_{{\boldsymbol\theta}_0} ( y_i|{\Y _{i-p}^{i-1}})}\nonumber\\
&=\sum_{i={j}}^k\underbrace{\left[\frac{1}{2}\log \frac{\sigma_\nu^2}{\sigma_u^2}-\frac{{\res_{i,{\boldsymbol\theta}_1}}^2}{2\sigma_u^2}+\frac{{\res_{i,{\boldsymbol\theta}_0}}^2}{2\sigma_\nu^2}\right]}_{s_i}.
\end{align}
Given a target false alarm period $\gamma$, the threshold is given by
$h \approx \ln (\gamma)$ \cite{Basseville}.
Note that at each time $k$, the test statistic $g_k$ is computed and compared with the threshold $h$. $T$ is the first time that $g_k$ exceeds $h$ and when the disturbance is declared to occur. If both ${\boldsymbol\theta}_0$ and ${\boldsymbol\theta}_1$ are exactly known, \eqref{general_CUSUM} is equivalent to the CUSUM test and the decision statistic $g_k$ can be recursively computed as
\begin{align}\label{CUSUM}
&g_k=\left(g_{k-1}+s_k\right)^+,
\end{align}
where $(x)^+ \triangleq  \max\{x, 0\}$. In essence, \eqref{CUSUM} forgets the past (i.e., resets) whenever $g_k\leq 0$, which relieves the detector from storing all previous observations in memory compared to that in \eqref{general_CUSUM}. To that end, \eqref{CUSUM} can also be represented in the following equivalent form \cite[Chapter 2]{Basseville}:
\begin{align}
&N_k=N_{k-1}\mathbbm{1}_{\{g_{k-1}>0\}}+1,\label{recursive_GLR}\\
&g_k=\left( S_{k-N_k+1}^k\right)^+, \label{recursive_CUSUM1}
\end{align}
where $N_k$ is number of observations at time $k$ since the last time of reset at time $k-N_k$ and $\mathbbm{1}_{\{\cdot\}}$ is the indicator function.

However, in the disturbance detection problem considered here, the post-change parameter ${\boldsymbol\theta}_1$ is unknown since many types of disturbances may potentially occur. Hence the CUSUM test cannot be directly applied here.
One solution is the weighted CUSUM, which requires a prior distribution on the unknown parameters and averages the decision statistic with respect to this presumed prior \cite[Chapter 2.4.2]{Basseville}. While the choice of the prior affects the performance substantially, there is no well justified prior available for the various power disturbance signals. Moreover, the computational complexity is usually high due to the multidimensional integral with respect to the prior density. The work \cite{Xingze10} adopts this method and assumes that the parameters are independent over time for reasonable computational complexity.
Yet another approach is the generalized log-likelihood ratio (GLR) test, where we substitute ${\boldsymbol\theta}_1$ with its  maximum-likelihood (ML) estimate  in \eqref{recursive_CUSUM1}:
\begin{align}
&g_k=\left( \sup_{{\boldsymbol\theta}_1} S_{k-N_k+1}^k\right)^+.\label{recursive_GLR1}
\end{align}
This is a desirable method when the statistical property of post-change parameters is not known a priori.
Still, the maximization in \eqref{recursive_GLR1} is difficult to solve since $S_{k-N_k+1}^k$ given by \eqref{LLR_N} is not a convex function of the parameters. Moreover, solving \eqref{recursive_GLR1} at every time $k$ leads to high computational complexity. In this paper, we apply a generalized local likelihood ratio (GLLR) test to solve our problem, which is elaborated in the next subsection.

\subsection{Generalized Local Likelihood Ratio Test}

The GLLR detector is based on the assumption that the parameter change in \eqref{def:hypothesis} is small, i.e., ${\boldsymbol\theta}_1\approx{\boldsymbol\theta}_0$. The corresponding test under this assumption is called the locally optimal test, meaning that the detector is asymptotically optimal as ${\boldsymbol\theta}_1 \rightarrow {\boldsymbol\theta_0}$. Since we have no prior knowledge of the disturbance, assuming that the change is small corresponds to the worst-case scenario that is most difficult to detect. On the other hand, if the disturbance induces significant changes, then it can be easily detected by any simple detection schemes.

The key idea of the local approach is to approximate the decision statistic  by a linear expansion around the nominal (pre-change) parameter. Thus we begin by expanding the conditional log-likelihood ratio in \eqref{LLR_N} up to the second order, yielding
\begin{align}\label{eq:s_approx_1}
S_j^k\approx \bar{S}_j^k &= \R^\intercal \left(\sum_{i=j}^k\left.\frac{\partial s_i}{\partial {\boldsymbol\theta}_1}\right|_{{\boldsymbol\theta}_1={\boldsymbol\theta}_0}\right)+\frac{1}{2}\R^\intercal \left(\sum_{i=j}^k\left.\frac{\partial^2 s_i}{\partial {\boldsymbol\theta}_1^2}\right|_{{\boldsymbol\theta}_1={\boldsymbol\theta}_0}\right)\R\nonumber\\&=\R^\intercal\left(\sum_{i=j}^k{\bf z}_i\right)-\frac{1}{2}\R^\intercal \left(\sum_{i=j}^k{\bf w}_i\right)\R,
\end{align}
where $\R\triangleq {\boldsymbol\theta}_1-{\boldsymbol\theta}_0$ and
\begin{align}
{\bf z}_i&\triangleq \left. \frac{\partial s_i}{\partial {\boldsymbol\theta}_1}\right|_{{\boldsymbol\theta}_1={\boldsymbol\theta}_0}\nonumber\\&=\left.\partial \left(-\frac{{\res_{i,{\boldsymbol\theta}_1}}^2}{2\sigma_u^2}-\frac{\log 2\pi \sigma_u^2}{2} \right)/\partial {\boldsymbol\theta}_1 \right|_{{\boldsymbol\theta}_1={\boldsymbol\theta}_0} \nonumber\\&=\left.\left[
\begin{array}{c}
\frac{1}{\sigma_u^2}{\res_{i,{\boldsymbol\theta}_1}} {\Y_{i-p}^{i-1}}\\
\frac{1}{\sigma_u}(\frac{{\res_{i,{\boldsymbol\theta}_1}}^2}{\sigma_u^2}-1)\\
\frac{{\res_{i,{\boldsymbol\theta}_1}}}{\sigma_u^2}
\end{array}
\right]\right|_{{\boldsymbol\theta}_1={\boldsymbol\theta}_0}=
\left[
\begin{array}{c}
\frac{1}{\sigma_\nu^2}y_i {\Y_{i-p}^{i-1}}\\
\frac{1}{\sigma_\nu}(\frac{{y_i}^2}{\sigma_\nu^2}-1)\\
\frac{{y_i}}{\sigma_\nu^2}
\end{array}\right],
\end{align}
\begin{align}
{\bf w}_i&\triangleq -\frac{\partial^2 s_i}{\partial {\boldsymbol\theta}_1^2}\left|\right._{{\boldsymbol\theta}_1={\boldsymbol\theta}_0}\nonumber\\&=\frac{1}{\sigma_u^2}\left.\left[
\begin{array}{ccc}
{\Y_{i-p}^{i-1}}{{\Y_{i-p}^{i-1}}}^\intercal & \frac{2\res_{i,{\boldsymbol\theta}_1}}{\sigma_u}{\Y_{i-p}^{i-1}} & {\Y_{i-p}^{i-1}}\\
\frac{2\res_{i,{\boldsymbol\theta}_1}}{\sigma_u}{{\Y_{i-p}^{i-1}}}^\intercal & \frac{3{\res_{i,{\boldsymbol\theta}_1}}^2}{\sigma_u^2}-1 & \frac{2\res_{i,{\boldsymbol\theta}_1}}{\sigma_u} \\
\Y_{i-p}^{i-1} & \frac{2\res_{i,{\boldsymbol\theta}_1}}{\sigma_u} & 1
\end{array}
\right]\right|_{{\boldsymbol\theta}_1={\boldsymbol\theta}_0}\nonumber\\&=\frac{1}{\sigma_\nu^2}
\left[
\begin{array}{ccc}
{\Y_{i-p}^{i-1}}{{\Y_{i-p}^{i-1}}}^\intercal & \frac{2y_{i}}{\sigma_\nu}{\Y_{i-p}^{i-1}}& \Y_{i-p}^{i-1}\\
\frac{2y_{i}}{\sigma_\nu}{{\Y_{i-p}^{i-1}}}^\intercal & \frac{3{y_{i}}^2}{\sigma_\nu^2}-1 & \frac{2y_i}{\sigma_\nu}\\
\Y_{i-p}^{i-1} & \frac{2y_i}{\sigma_\nu} & 1
\end{array}
\right].
\end{align}

We still need to decide the change direction $\R$. Note that \eqref{eq:s_approx_1} is a quadratic function of the change direction $\R$, which implies that after we replace $S^k_{k-N_k+1}$ in \eqref{recursive_GLR1} with its second-order approximation $\bar{S}_{k-N_k+1}^k$, i.e., $\sup_{\boldsymbol\theta_1} S_{k-N_k+1}^k\approx \sup_{\bf r} \bar{S}^k_{k-N_k+1}$, then the optimization problem can be solved analytically. We rewrite \eqref{eq:s_approx_1} as
\begin{align}\label{eq:s_approx_2}
\bar{S}_j^k = &\; \R^\intercal\left(\sum_{i=j}^k{{\bf z}}_i\right) -\frac{(k-j+1)}{2}\R^\intercal \left(\sum_{i=j}^k\frac{{{\bf w}}_i}{(k-j+1)}\right)\R\nonumber\\\to&\; \R^\intercal\left(\sum_{i=j}^k{\bf z}_i\right)-\frac{k-j+1}{2}\R^\intercal {\bf J}({\boldsymbol\theta}_0)\R,
\end{align}
as $k-j+1 \to \infty$, corresponding to the case of large number of samples. This is obtained by considering
\begin{align}
\sum_{i=j}^k\frac{{{\bf w}}_i}{k-j+1}\to \E_{\boldsymbol\theta}({\bf w}_i),
\end{align}
which can be approximated by the local assumption that ${\boldsymbol\theta}_0\approx {\boldsymbol\theta}_1$, and hence

\ignore{as $(k-j+1)\to \infty$, corresponding to the case of either large number of samples, and/or large number of sensors. This is obtained by considering
\begin{align}
\sum_{i=j}^k\frac{{{\bf w}}_i}{k-j+1}\to \E_{\boldsymbol\theta}({\bf w}_i),
\end{align}
which can be approximated by the local assumption that ${\boldsymbol\theta}_0\approx {\boldsymbol\theta}_1$, and hence}
\begin{align}\label{FisherInf}
\E_{\boldsymbol\theta}({\bf w}_i)&\approx \E_{{\boldsymbol\theta}_0}({\bf w}_i)=
\left[
\begin{array}{ccc}
{\bf I} & {\bf 0} & {\bf 0}\\ {\bf 0} & \frac{2}{\sigma_\nu^2} & 0\\
{\bf 0} & 0 & \frac{1}{\sigma_\nu^2}
\end{array}
\right],
\end{align}
where ${\bf I}$ is the $p\times p$ identity matrix. We denote the matrix in  \eqref{FisherInf} by ${\bf J}({\boldsymbol\theta}_0)$ because it is, in fact, the Fisher information of the parameter  ${\boldsymbol\theta}_0$, i.e.,
\begin{align}
{\bf J}({\boldsymbol\theta}_0)\triangleq \E\left[\left.-\frac{\partial^2 \log\! f_{{\boldsymbol\theta}}}{\partial\; {\boldsymbol\theta}^2}\right|_{{\boldsymbol\theta}={\boldsymbol\theta}_0}\right].
\end{align}
We now constrain $\R$ to be on a small ellipse, i.e., $\R^\intercal {\bf J}({\boldsymbol\theta}_0)\R=b^2$. Based on \eqref{eq:s_approx_2}, we can then compute the change direction analytically as follows,
\begin{align}\label{eq:solve_direction}
\tilde{S}_{j}^k\triangleq \sup_{\R^\intercal {\bf J}({\boldsymbol\theta}_0)\R=b^2} \bar{S}_{j}^k=\sup_{\R^\intercal {\bf J}({\boldsymbol\theta}_0)\R=b^2}\R^\intercal\left({\sum_{i=j}^k{\bf z}_i}\right)-\frac{k-j+1}{2}\R^\intercal {\bf J}({\boldsymbol\theta}_0)\R.
\end{align}
Decomposing ${\bf J}({\boldsymbol\theta}_0)={\bf B^\intercal B}$ with
\begin{align}
{\bf B}\triangleq \left[
\begin{array}{ccc}
{\bf I} & {\bf 0} & {\bf 0}\\
{\bf 0} & \frac{\sqrt{2}}{\sigma_\nu} & 0\\
{\bf 0} & 0& \frac{1}{\sigma_v}
\end{array}\right],
\end{align}
it follows that $\R^\intercal {\bf J}({\boldsymbol\theta}_0)\R=({\bf B}\R)^\intercal ({\bf B}\R)=\tilde{\R}^\intercal\tilde{\R}=b^2$.
Then \eqref{eq:solve_direction} becomes
\begin{align}\label{eq:central_S}
\tilde{S}_j^k=\sup_{\tilde{\R}^\intercal \tilde{\R}=b^2} \bar{S}_j^k&=\sup_{\tilde{\R}^\intercal \tilde{\R}=b^2}\tilde{\R}^\intercal({\bf B}^\intercal)^{-1}\left(\sum_{i=j}^k{{\bf z}}_i\right)-\frac{k-j+1}{2}b^2\nonumber\\&=\sup_{\tilde{\R}^\intercal \tilde{\R}=b^2}\tilde{\R}^\intercal\left(\sum_{i=j}^k\tilde{{\bf z}}_i\right)-\frac{k-j+1}{2}b^2\nonumber\\
&=bU_j^k-\frac{k-j+1}{2}b^2,
\end{align}
with
\begin{align}
U_j^k\triangleq \lVert\sum_{i=j}^k\tilde{\bf z}_i\rVert \qquad
\text{and}\qquad
\tilde{\bf z}_i\triangleq ({\bf B^\intercal})^{-1}{\bf z}_i=\left[
\begin{array}{c}
\frac{1}{\sigma_\nu^2}{y_i} {\Y_{i-p}^{i-1}}\\
\frac{1}{\sqrt{2}}(\frac{{y_i}^2}{\sigma_\nu^2}-1)\\
\frac{y_i}{\sigma_v}
\end{array}\right]. \label{eq:adjusted_z}
\end{align}
Note that the statistic $\tilde{\bf z}_i$ is a function of the observations and intuitively relates to the disturbance model given by \eqref{model}. In specific, the term of $y_i\Y_{i-p}^{i-1}/\sigma_\nu^2$ accounts for the autocorrelation of meter samples, $\left(\frac{y_i^2}{\sigma_\nu^2}-1\right)/\sqrt{2}$ captures the variance increase of meter samples and $y_i/\sigma_\nu$ monitors the mean shift.

Finally, incorporating the generalized local likelihood ratio statistics \eqref{eq:central_S}-\eqref{eq:adjusted_z} into the sequential change detector \eqref{recursive_GLR}-\eqref{recursive_CUSUM1}, we summarize the GLLR detector for voltage disturbance detection as
\begin{align}
&N_k=N_{k-1}\mathbbm{1}_{\{\tilde{g}_{k-1}>0\}}+1,\label{recursive_LGLR}\\
&\tilde{g}_k=\left( \tilde{S}_{k-N_k+1}^k\right)^+,\label{recursive_LGLR_1}\\
&\tilde{T}=\inf \{k: \tilde{g}_k \geq h\}. \label{recursive_LGLR_2}
\end{align}

We highlight the following features of the above disturbance detector:
\begin{itemize}
\item Benefiting from the ML estimate of the unknown parameters, it can track the  disturbance adaptively, thus is robust to a variety of distortions.
\item The decision statistic is an analytical function of the observations given by \eqref{eq:central_S}-\eqref{eq:adjusted_z}, which is derived based on the  assumption of small $\R$, thus is easy to compute. Moreover, similar to the CUSUM test, it has a recursive form and thus can be efficiently implemented.
\item It is implemented by sequentially observing the meter readings, thus is expected to outperform the RMS method and the STFT method in terms of detection delay, which rely on a sampling window, within which the RMS and spectrum are evaluated respectively.
\end{itemize}
In Section IV, we provide extensive experiments based on the realistic disturbance signals to demonstrate the superior performance of the proposed GLLR detector compared to the existing methods.
\section{Cooperative Monitoring Based on Level-Triggered Sampling}

In this section, we consider the cooperative detection of disturbances in an area where the buses are exposed to power signal disturbance at the same time. Cooperative detection takes advantage of the fact that a number of electrical buses can be affected at the same by a power quality event, and allows us to achieve faster decision. Particularly, it is implemented by deploying multiple meters across the network that communicate wirelessly with a central meter which is responsible for monitoring the power quality. Consider $L$ meters that are linked wirelessly with a central meter and perform the cooperative disturbance detection. The straightforward scheme is to make the distributed measurements fully available to the central meter by transmitting very finely (infinite-bit) quantized measurements at every sampling instant, i.e., the centralized setup. However, in practice,  the wireless links between the distributed meters and the central meter are characterized by limited bandwidth. Therefore, in designing a practical system, two constraints need to be considered, namely the {\it rate constraint} (i.e., the distributed meters should communicate with the central meter at a lower rate than the local sampling rate) and the {\it quantization constraint} (i.e., each meter should transmit a small number of bits every time it communicates with the central meter). In particular, considering the
high sampling rate at distributed meters (e.g., for $60 \text{Hz}$ AC supply in North America, the sampling rate could be $3.6\text{kHz}$ with $64$ samples per cycle) and the large number of quantization bits in order to achieve an acceptable  accuracy at the central meter, it is inefficient to inform the central meter of the local observations at every sampling instant. Thus the decentralized detection, where the distributed meters communicate with the central meter in some low-rate fashion, becomes necessary.
In this section, we propose a level-triggered sampling scheme which efficiently lowers the communication overhead in terms of both the communication frequency and the number of information bits at each transmission, while preserving the time resolution of the disturbance detection.

We begin with deriving the centralized multi-meter GLLR detection scheme. The pre-change and post-change signal model in the multi-meter setup is written as
\begin{align}
\left\{\begin{array}{ll}
y_t^{(\ell)}=\nu_t^{(\ell)} & t<t_0,\\
y_t^{(\ell)}=\mu^{(\ell)}+\sum_{j=1}^pa_j^{(\ell)}y^{(\ell)}_{t-j}+u^{(\ell)}_t, &  t\ge t_0,
\end{array}\right.\qquad \ell=1, 2, \ldots, L.
\end{align}
Here we assume that ${\boldsymbol\theta}^{(\ell)}=[a_1^{(\ell)}, \ldots, a_p^{(\ell)}, \mu^{(\ell)}, \sigma_u^{(\ell)}]$ varies with $\ell$, because, in general, the distributed meters observe power voltage at different buses, thus the resulting disturbance signals are not necessarily the same. The driving process $u_t^{(\ell)}$ are assumed to be independent across meters. Correspondingly, the log-likelihood ratio function in \eqref{LLR_N} becomes
\begin{align}
S_{j}^k
&\triangleq
\sum_{\ell=1}^L\underbrace{\sum_{i={j}}^k\left[\frac{1}{2}\log \frac{\sigma_\nu^2}{\sigma_u^2}-\frac{{\res_{i,{\boldsymbol\theta}_1}^{(\ell)}}^2}{2\sigma_u^2}+\frac{{\res_{i,{\boldsymbol\theta}_0}^{(\ell)}}^2}{2\sigma_\nu^2}\right]}_{{S_j^k}^{(\ell)}},
\end{align}
where $\varepsilon_{t,{\boldsymbol\theta}_0}^{(\ell)}\triangleq y_t^{(\ell)}$ and $\varepsilon_{t,{\boldsymbol\theta}_1}^{(\ell)}\triangleq y_t^{(\ell)}-\mu^{(\ell)}-\sum_{j=1}^pa^{(\ell)}_jy^{(\ell)}_{t-j}$. Moreover, the generalized log-likelihood ratio in \eqref{recursive_GLR1} is evaluated as
\begin{align}\label{GLR_M}
\tilde{S}_j^k\triangleq \sup_{\boldsymbol\theta_1^{(\ell)}, \ell=1, \ldots, L} S_j^k= \sum_{\ell=1}^L \left(\sup_{\boldsymbol\theta_1^{(\ell)}} {S_j^k}^{(\ell)}\right).
\end{align}
Applying the same local approximation as in the last section to $\sup_{\boldsymbol\theta_1^{(\ell)}} {S_j^k}^{(\ell)}$, then we further evaluate \eqref{GLR_M} as follows:
\begin{align}
\tilde{S}_j^k=
\sum_{\ell=1}^L\underbrace{\left(b{U_j^k}^{(\ell)}-\frac{k-j+1}{2}b^2\right)}_{\tilde{S}_j^{k, (\ell)}},\label{eq:decentral_S}
\end{align}
where
\begin{align}
{U_j^k}^{(\ell)}\triangleq \lVert\sum_{i=j}^k\tilde{\bf z}_i^{(\ell)}\rVert^2, \qquad \tilde{\bf z}_i^{(\ell)}\triangleq \left[
\begin{array}{c}
\frac{1}{\sigma_\nu^2}{y_i^{(\ell)}} {\left(\Y^{(\ell)}\right)_{i-p}^{i-1}}\\
\frac{1}{\sqrt{2}}(\frac{{y_i^{(\ell)}}^2}{\sigma_\nu^2}-1)\\
\frac{y_i^{(\ell)}}{\sigma_v}
\end{array}\right],\label{eq:individual}
\end{align}
and ${\left(\Y^{(\ell)}\right)_j^k}\triangleq [y^{(\ell)}_k, y^{(\ell)}_{k-1}, \ldots, y^{(\ell)}_{j}]^\intercal$ denote the observations from sample $k$ to $j$ at sensor $\ell$. As a result, the cooperative multi-meter GLLR detector is summarized as
\begin{align}
&N_k=N_{k-1}\mathbbm{1}_{\{\tilde{g}_{k-1}>0\}}+1,\label{M1}\\
&\tilde{g}_k=\left(\sum_{\ell=1}^L \tilde{S}_{k-N_k+1}^{k, (\ell)}\right)^+,\\
&\tilde{T}=\inf \{k: \tilde{g}_k \geq h\}.\label{M3}
\end{align}
Comparing \eqref{M1}-\eqref{M3} with \eqref{recursive_LGLR}-\eqref{recursive_LGLR_2}, we find that the cooperative GLLR detector differs from the single-meter GLLR detector by summing distributed $\tilde{S}_{k-N_k+1}^{k, (\ell)}$ instead of only using that at one meter. As such, the centralized GLLR detector requires the distributed meters to quantize and transmit the local statistic $\tilde{S}_{k-N_k+1}^{k, (\ell)}$ to the central meter at every local sampling instant $k$.

We next consider the decentralized implementation of \eqref{M1}-\eqref{M3}. In the decentralized setup, it is important to devise an efficient communication scheme between the distributed meters and the central meter, by which the local statistics are sent to the central meter less frequently and using small number of bits at each transmission. In the following subsections, we propose an efficient decentralized implementation based on the level-triggered communication scheme at the sensors and its associated decision rule at the central meter.

\subsection{Decentralized Detection Based on Level-Triggered Sampling}

We first describe the level-triggered sampling strategy, which is essentially a single-bit quantization of the local statistic, where the transmission of the local statistic is only triggered once it hits a certain value, thus is observation-adaptive. Moreover, all meters communicate with the central meter asynchronously, which avoids the use of a global clock for synchronization.

We simplify the notation of the local test statistic at the $\ell$th meter $\{{\tilde{S}_{k-N_k+1}^{k, {(\ell)}}}\}$ as $\{\tilde{S}_k^{(\ell)}\}$, because $N_k$ is uniquely determined by $k$, and denote the $n$th communicating time of the $\ell$th meter as $k_n^{\ell}$.
Note that at the $\ell$th meter, we can decompose the test statistic as
\begin{align}\label{eq:partition}
\tilde{S}^{(\ell)}_{k}=\tilde{S}^{(\ell)}_{k}-\tilde{S}^{(\ell)}_{k^\ell_{n-1}}+\tilde{S}^{(\ell)}_{k^\ell_{n-1}}-\ldots-\tilde{S}^{(\ell)}_{k_1^\ell}+\tilde{S}^{(\ell)}_{k_1^\ell}-\tilde{S}^{(\ell)}_0,
\end{align}
where $\tilde{S}^{(\ell)}_0=0$, and $k_n^{\ell}$ is recursively defined as
\begin{align}\label{def:tn}
k_n^{\ell}\triangleq \inf \left\{k>k_{n-1}^{\ell}: {\tilde{S}_k}^{(\ell)}-{\tilde{S}_{k_{n-1}}}^{(\ell)}\notin (-\underline{\Delta}, \overline{\Delta})\right\}, \quad k_0^{\ell}=0, \; \tilde{S}_0^{(\ell)}=0,
\end{align}
where $\uD$ and $\lD$ are positive constants, selected to control the frequency of transmission and known to the central meter. According to \eqref{def:tn}, each meter informs the central meter of its local statistic every time it cumulates to exit the interval $[-\lD,\uD]$. In the ideal case, $\tilde{S}_{k_n^\ell}^{(\ell)}-\tilde{S}_{k_{n-1}^\ell}^{(\ell)}$ hits the boundary exactly in \eqref{def:tn}, i.e., ${\tilde{S}_{k_n^\ell}}^{(\ell)}-{\tilde{S}_{k_{n-1}^\ell}}^{(\ell)}= -\underline{\Delta} \;\text{or} \;\overline{\Delta}$. Then the local statistic can be delivered by sending only one-bit information of which boundary is hit to the central meter.
In particular, the $n$th one-bit message transmitted by the $\ell$th meter  is given by
\begin{align}\label{1bit-inf}
  x_{n}^{(\ell)}=\left\{
  \begin{array}{cc}
  1, &\quad \text{if} \;\; {\tilde{S}_{k_n^\ell}}^{(\ell)}-{\tilde{S}_{k^\ell_{n-1}}}^{(\ell)}\geq \uD,\\
  -1, &\quad \text{if} \;\; {\tilde{S}_{k_n^\ell}}^{(\ell)}-{\tilde{S}_{k^\ell_{n-1}}}^{(\ell)}\leq -\lD.
  \end{array}\right.
\end{align}
In essence, the central meter uniformly samples the local statistic in its value domain instead of in the time domain to lower the transmission frequency.
Moreover, the quantization of local statistic is no longer needed, which substantially decreases the amount of data at each transmission.  The level-triggered sampling scheme at each meter is summarized as Algorithm 1.
 Note that the reset signal in the procedure corresponds to the indicator function in the GLLR test \eqref{M1}: recalling that $N_k$ is the number of observations for computing the local statistic, when the global statistic at the central meter $\tilde{S}_k\leq 0$, a reset signal is broadcast to all meters informing them to reset $N_k=1$; otherwise, with no reset signal, $N_k$ keeps increasing.
\begin{algorithm}
\caption{\bf : Level-triggered sampling of the GLLR statistic at the $\ell$th meter}
\begin{algorithmic}[1]
\STATE Initialization: $k\leftarrow 0$
\STATE Reset: $\lambda\leftarrow 0,  N \leftarrow 1$%, compute $\tilde{S}_k^{k, (\ell)}$ ($\tilde{S}_0^{0, (\ell)}=0$)
\STATE {\bf while} $\tilde{S}_{k-N+1}^{k, (\ell)}-\lambda\in (-\underline{\Delta},\overline{\Delta})$ {\bf do}
\STATE \quad $k\leftarrow k+1$
\STATE \quad Check the reset signal broadcasted by the central meter:
\STATE \quad {\bf if} present {\bf then}\\
%\STATE \qquad $N\leftarrow1$, $\lambda \leftarrow 0$\\
\STATE \qquad go to line 2\\
\STATE \quad {\bf else}\\\STATE\qquad $N\leftarrow N+1$\\
\STATE \quad {\bf end if}
\STATE \quad Compute $\tilde{S}_{k-N+1}^{k, (\ell)}$ by \eqref{eq:decentral_S}-\eqref{eq:individual}
\STATE {\bf end while}
\STATE Send $x_k^{(\ell)}=\text{sign}(\tilde{S}_{k-N+1}^{k, (\ell)}-\lambda)$ to the central meter
\STATE $\lambda\leftarrow \tilde{S}_{k-N+1}^{k, (\ell)}$
\STATE Check the reset signal broadcast by the central meter:
\STATE \quad {\bf if} present {\bf then} go to line 2\\
\STATE \quad {\bf else} go to line 3.
\end{algorithmic}
\end{algorithm}
The above transmission scheme features an inherent data compression and adaptive communication between the local meters and the central meter. Moreover, the one-bit transmission induces significant savings in bandwidth and transmission power.

On the other side, the central meter receives the information bits from each meter asynchronously and updates the global running statistic as follows:
\begin{align}
\tilde{S}_{k}&=\tilde{S}_{k-1}+\sum_{\ell=1}^L\left(\mathbbm{1}_{\{k=k_n^{\ell}, \;x_n^{(\ell)}=1\}}\overline{\Delta}-\mathbbm{1}_{\{k=k_n^{\ell}, \;x_n^{(\ell)}=-1\}}\underline{\Delta}\right)\nonumber\\&=\sum_{\ell=1}^L\sum_{n:\;k_n^{\ell}<k}\left(\mathbbm{1}_{\{
x_n^{(\ell)}=1\}}\overline{\Delta}-\mathbbm{1}_{\{x_n^{(\ell)}=-1\}}\underline{\Delta}\right),
\end{align}
which is essentially the decentralized counterpart of \eqref{eq:decentral_S}.
Every time the global statistic is updated at the central meter, it is used to perform the GLLR test given by \eqref{M1}-\eqref{M3}. There are two decisions to make, i.e., triggering the alarm that a disturbance is detected or continuing to receive information bits from the meters. The procedure at the central meter is summarized as Algorithm 2.

\begin{algorithm}
\caption{\bf : Operations at the central meter}
\begin{algorithmic}[1]
\STATE Initialization: $\tilde{S}\leftarrow 0, k \leftarrow 0$
\STATE {\bf while} $\tilde{S}< h$ {\bf do}
\STATE \quad $k\leftarrow k+1$
\STATE \quad Listen to the meters and receive information bits, say, $r_1$ ``$+1$''s and $r_2$ ``$-1$''s
\STATE \quad $\tilde{S}\leftarrow\tilde{S}+r_1\overline{\Delta}-r_2\underline{\Delta}$
\STATE \quad {\bf if} $r_1+r_2\neq 0$ \& $\tilde{S}\leq 0$ {\bf then}\\
\STATE \qquad $\tilde{S}\leftarrow 0$
\STATE \qquad broadcast the reset signal to all meters \\
\STATE \quad {\bf end if}
\STATE {\bf end while}
\STATE Trigger the disturbance alarm and broadcast the reset signal to all meters
\end{algorithmic}
\end{algorithm}

Through the level-triggered sampling scheme, we efficiently recover the decision statistic at the central meter by collecting local statistics from meters. Specifically, compared to the centralized setup where observations are transmitted at every sampling instant with multiple quantization bits, the level-triggered sampling features lower communication frequency (which can be controlled by the parameters $\lD$ and $\uD$) and one-bit representation of each sample.

\subsection{Enhancement}

\begin{figure}
\centering
\subfigure[Original]{
\includegraphics[width=0.48\textwidth]{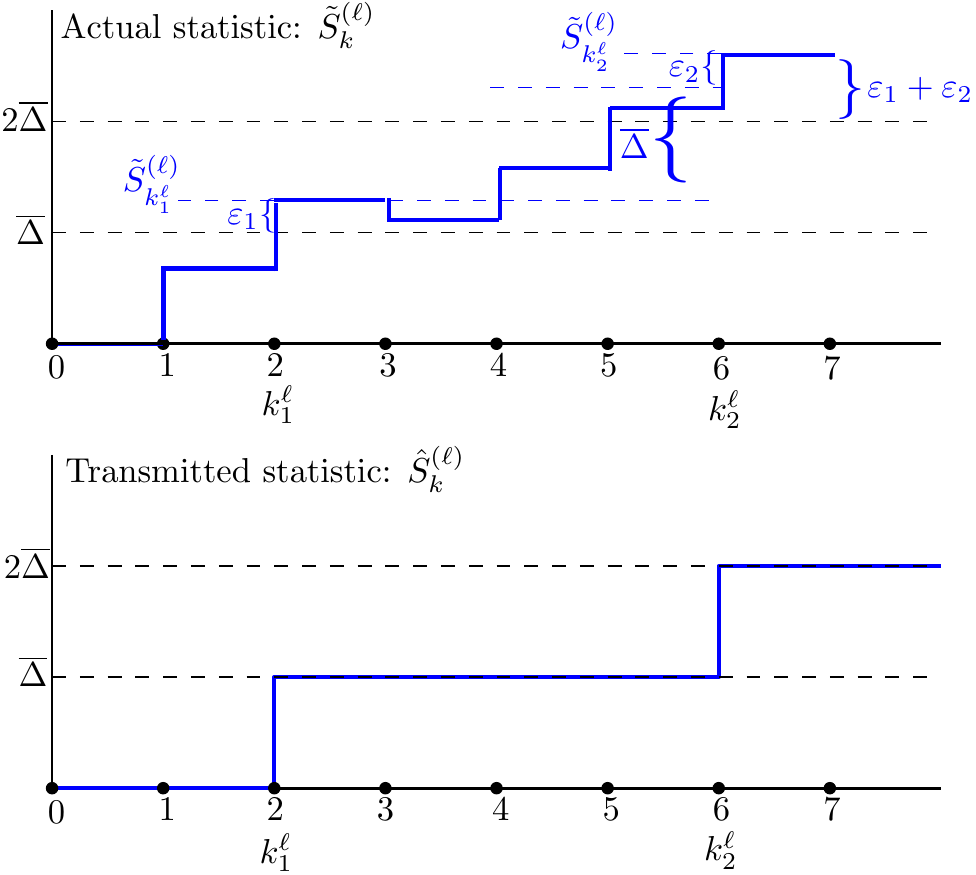}}
\subfigure[Enhanced]{
\includegraphics[width=0.48\textwidth]{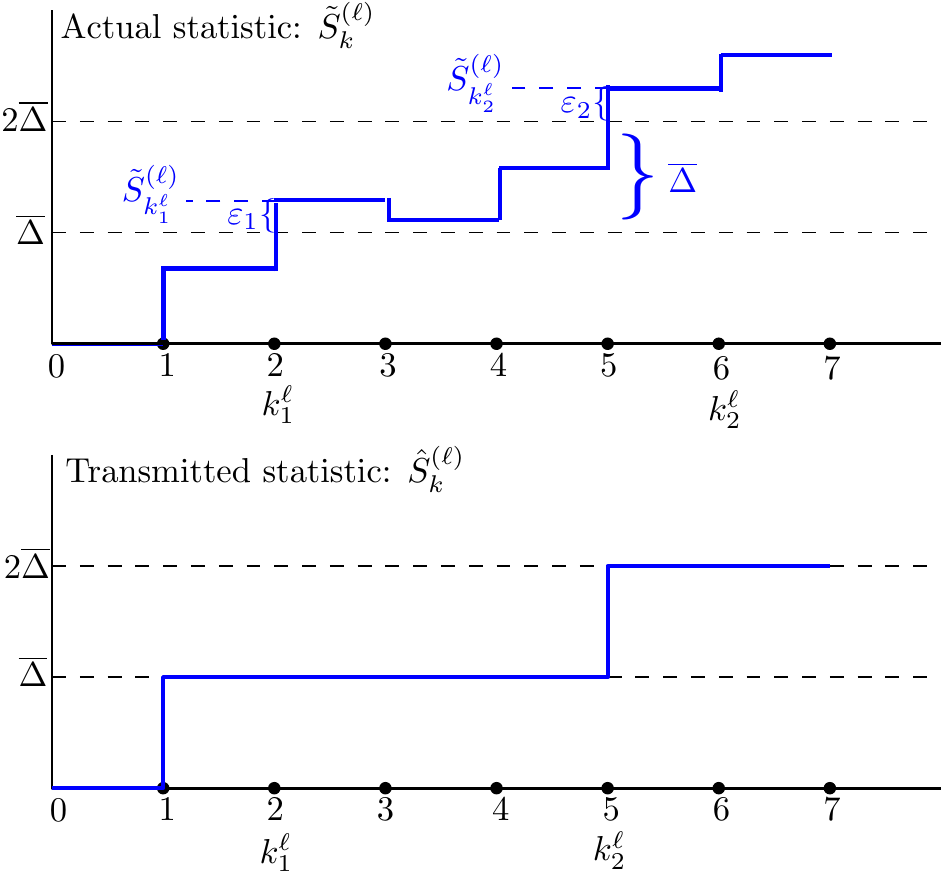}}
\caption{Illustration of the original and enhanced level-triggered sampling in the presence of overshoot errors.}\label{fig:overshoot_illustrate}
\end{figure}

In this subsection, we consider the realistic case where the local statistics do not exactly hit the local thresholds at each level-triggered sampling instant. Under such circumstance,  
%Then, when the increment $\tilde{S}^{(\ell)}_{k_n^\ell}-\tilde{S}^{(\ell)}_{k^\ell_{n-1}}$ exceeds the $[-\lD, \uD]$ interval, it does not exactly hit the boundaries.
%varies in the range $[-\infty,-\underline{\Delta})$ and $[\overline{\Delta}, \infty)$, i.e., $\tilde{S}^{(\ell)}_{k^\ell_{n}}-\tilde{S}^{(\ell)}_{k_{n-1}^\ell}$ could overhit $[-\lD,\uD]$ by arbitrary value.
information loss will incur due to the overshoot error.  For example, in Fig. \ref{fig:overshoot_illustrate}(a), when the increment of the actual statistic $\tilde{S}_{k}^{(\ell)}$ at the $\ell$th meter first exceeds the upper threshold $\uD$ at time $k_1^\ell$, an error $\varepsilon_1\triangleq \tilde{S}_{k_1^\ell}^{(\ell)}-\tilde{S}_{0}^{(\ell)}-\uD$ is incurred, where $\tilde{S}_{0}^{(\ell)}=0$. The next level-triggered sampling occurs at time $k_2^\ell$ when $\tilde{S}_{k_2^\ell}^{(\ell)}-\tilde{S}_{k_1^\ell}^{(\ell)}> \uD$ and again an error $\varepsilon_2\triangleq \tilde{S}_{k_2^\ell}^{(\ell)}-\tilde{S}_{k_1^\ell}^{(\ell)}-\uD$ is incurred, ending up with an overall error $\tilde{S}_{k_2^\ell}^{(\ell)}-\hat{S}_{k_2^\ell}^{(\ell)}=\varepsilon_1+\varepsilon_2$, where $\hat{S}_{k}^{(\ell)}$ is the transmitted statistic up to time $k$ by meter $\ell$ given by \eqref{eq:trans_S}.
%Two issues may arise: the {\it overshoot} and the {\it saturation}. Both issues originate from the discontinuity of the local LLR process, due to which the statistic at transmission time is not necessarily equal to the upper or lower boundary, thus introducing information loss. The difference between the overshoot and the boundary saturation is that overshoot occurs when waiting for another observation results in exiting the boundary, while the saturation exists when every single observation results in a much larger increment of LLR than the local thresholds, such that adding the threshold values at the central meter severely underestimate the local observations. We are going to propose different methods to overcome these two issues.
A main problem with the level-triggered sampling scheme in \eqref{def:tn} is that the overshoot errors accumulate over time. In general, using \eqref{eq:partition}-\eqref{def:tn}, we can write the actual statistic at the $\ell$th meter as
\begin{align}\label{eq:err_cu}
\tilde{S}^{(\ell)}_{k}=(\Delta_n+\varepsilon_n)+(\Delta_{n-1}+\varepsilon_{n-1})+\cdots+(\Delta_1+\varepsilon_1),\qquad k_n^\ell \leq k< k_{n+1}^\ell\;,
\end{align}
and the corresponding transmitted statistic
\begin{align}\label{eq:trans_S}
\hat{S}_{k}^{(\ell)}=\Delta_n+\Delta_{n-1}+\cdots+\Delta_1,\qquad k_n^\ell \leq k< k_{n+1}^\ell\;,
\end{align}
where $\Delta_i\in \{\overline{\Delta} ,-\underline{\Delta}\}$ and $\varepsilon_i$ is the error incurred at the level-triggered sampling instant $k_i^\ell$. Then we have the overall error $\tilde{S}_{k_n^\ell}^{(\ell)}-\hat{S}_{k_n^\ell}^{(\ell)}=\varepsilon_1+\varepsilon_2+\cdots+\varepsilon_n$ after $n$ level-triggered samplings, resulting in significant distortion on the reconstructed global statistic at the central meter.

This problem is addressed in \cite{Yasin12} by introducing extra bits to quantize the overshoot. Here we propose a new simple but effective method to mitigate the overshoot error accumulation problem while preserving the single-bit transmission feature. In particular, we modify the stopping time \eqref{def:tn} to
\begin{align}\label{modified_overshoot}
k_n^{\ell}\triangleq \inf \left\{k>k_{n-1}^{\ell}: {\tilde{S}_k}^{(\ell)}-{\tilde{S}_{k^\ell_{n-1}}}^{(\ell)}+\varepsilon_{n-1}\notin (-\underline{\Delta}, \overline{\Delta})\right\}.
\end{align}
Under this strategy, at sampling instant $k_n^\ell$, the previous overshoot error $\varepsilon_{n-1}$ is incorporated in the current sampling. In this way, we have only the current overshoot error present at any time but there is no longer error accumulation. Fig. \ref{fig:overshoot_illustrate}(b) illustrates the modification in \eqref{modified_overshoot}. The same as in Fig. \ref{fig:overshoot_illustrate}(a), error $\varepsilon_1$ is incurred at the first level-triggered sampling instant $k_1^\ell$. However, the next sampling occurs at $k_2^\ell$ when $\tilde{S}_{k_2^\ell}^{(\ell)}-\tilde{S}_{k_1^\ell}^{(\ell)}+\varepsilon_1>\uD$. Since $\varepsilon_1$ is included in the transmitted value $\uD$, the overall error after two samplings $\tilde{S}_{k_2^\ell}^{(\ell)}-\hat{S}_{k_2^\ell}^{(\ell)}$ is now only $\varepsilon_2$. Similarly, $\varepsilon_2$ will be incorporated in the next level-triggered sampling while $\varepsilon_3$ occurs and so on. Simply put, the modification can be interpreted as that, instead of transmitting the increment between the current statistic and the statistic at the last level-triggered sampling instant, we transmit the increment between the current statistic and the total transmitted statistics up to the last sampling instant.

To incorporate \eqref{modified_overshoot} into Algorithm 1, the only necessary change is on Line 14, which is replaced by
\begin{align*}
\lambda\leftarrow \lambda+\mathbbm{1}_{\{x_n^{(\ell)}=1\}}\uD-\mathbbm{1}_{\{x_n^{(\ell)}=-1\}}\lD\;;
\end{align*}
whereas no change is needed for the operations at the central meter, i.e., Algorithm 2 remains the same.
\ignore{
\begin{figure}
\centering
\subfigure[$\uD=\lD=5$]{
\includegraphics[width=0.48\textwidth]{Figures/Enhanced_overshoot_55.eps}\label{fig:overshoot1}
}
\subfigure[$\uD=\lD=8$]{
\includegraphics[width=0.48\textwidth]{Figures/Enhanced_overshoot_88.eps}\label{fig:overshoot2}}
\caption{Comparison of the centralized and decentralized (original and enhanced)  statistics.}
\end{figure}
}
\ignore{
\subsection{Parameter Tuning}
To choose a proper $b$, first we introduce the detectability condition for the general sequential change detection \cite{Basseville}, which states that the two distributions before and after the occurrence of the disturbance are detectable with finite stopping time if and only if $\E_{{\boldsymbol\theta}_0}(\phi_k)<0<\E_{{\boldsymbol\theta}_1}(\phi_k)$, where $\phi_k$ represents the $k$th sample statistic. This condition implies that the test statistic takes different drifting directions between the pre-change and post-change distributions. Therefore, we choose $b$ such that the detectability condition holds, i.e.,
\begin{align}\label{detectability}
\E_{{\boldsymbol\theta}_0}(\tilde{s}_i)<0<\E_{{\boldsymbol\theta}_1}(\tilde{s}_i).
\end{align}
We begin with satisfying the first inequality. Note that
\begin{align}\label{detectability0}
\E_{{\boldsymbol\theta}_0}(\tilde{s}_i)=b\; \E_{{\boldsymbol\theta}_0}(\lVert\tilde{\bf z}_i\rVert)-\frac{b^2}{2} < b\sqrt{\E_{{\boldsymbol\theta}_0}(\lVert\tilde{\bf z}_i\rVert^2)}-\frac{b^2}{2}.
\end{align}
Thus it suffices to set $b$ such that the right-hand side of the inequality in \eqref{detectability0} to be smaller than zero, which is equivalent to setting
\begin{align}
b&\ge 2\sqrt{\E_{{\boldsymbol\theta}_0}(\lVert\tilde{\bf z}_i\rVert^2)}\nonumber\\
&=2 \left[\E_{{\boldsymbol\theta}_0}\left(\sum_{j=1}^p (y_iy_{i-j})^2/\sigma_v^4+\left(y_i^2/\sigma_v^2-1\right)^2/2+y_i^2/\sigma_v^2\right)\right]^{\frac{1}{2}}\nonumber\\&=2 \left[\sum_{j=1}^p \E_{{\boldsymbol\theta}_0}\left[(y_iy_{i-j})^2\right]/\sigma_v^4+\E_{{\boldsymbol\theta}_0}\left[\left(y_i^2/\sigma_v^2-1\right)^2\right]/2+\E_{{\boldsymbol\theta}_0}\left(y_i^2/\sigma_v^2\right)\right]^{\frac{1}{2}}\nonumber\\&=2\sqrt{0+1+1}=2\sqrt{2},
\end{align}
where the last inequality is obtained by noting that $y_i$ is white Gaussian noise  under ${\boldsymbol\theta}_0$, i.e., before the occurrence of disturbance. On the other hand, $\ell_2$ norm is a convex function, thus by Jensen's inequality, we have
\begin{align}\label{detectability1}
\E_{{\boldsymbol\theta}_1}(\tilde{s}_i)=b \;\E_{{\boldsymbol\theta}_1}(\lVert\tilde{\bf z}_i\rVert)-\frac{b^2}{2} > b \lVert\E_{{\boldsymbol\theta}_1}\tilde{\bf z}_i\rVert-\frac{b^2}{2},
\end{align}
and the detectability is guaranteed by setting the right-hand side of the inequality in \eqref{detectability1} to be greater than zero. To this end, we need to evaluate
\begin{align}
\lVert\E_{{\boldsymbol\theta}_1}\tilde{\bf z}_i\rVert=\left[\sum_{j=1}^p\left(\E_{{\boldsymbol\theta}_1}\left(y_iy_{i-j}\right)\right)^2+\left(\E_{{\boldsymbol\theta}_1}(y_i^2)/\sigma_v^2-1\right)^2/{2}+\left(\E_{{\boldsymbol\theta}_1}(y_i)\right)^2/\sigma_v^2\right]^{\frac{1}{2}}
\end{align}
in terms of signal parameter ${\boldsymbol\theta}_1$. We first introduce the shift operator $B: B^iy_t=y_{t-i}$, thus the disturbance \eqref{model}-\eqref{change_model} can be rewritten as
\begin{align}
\phi(B)y_t=\mu_t+\phi(B)v_t+w_t, \quad \text{and}\quad \phi(B)\triangleq 1-\sum_{j=1}^p a_j^tB^j,
\end{align}
which leads to
\begin{align}
y_t&=\phi(B)^{-1} \mu_t+\phi(B)^{-1}w_t+v_t\nonumber\\
&=\sum_{j=0}^\infty c_j\mu_{t-j}+\sum_{j=0}^\infty c_jw_{t-j}+v_t.
\end{align}
where $c_j$ can be computed recursively as
\begin{align}
c_0=1, \quad c_j=\sum_{k=1}^{\min\{p, j\}}a_kc_{j-k}, \quad j=1, 2, \ldots
\end{align}
\begin{align}
\E_{{\boldsymbol\theta}_1}\left(y_ty_{t-j}\right)\ge \sum_{k=0}^\infty c_k\mu_{t-k}\sum_{k=0}^\infty c_k\mu_{t-k-j}\\
\E_{{\boldsymbol\theta}_1}(y_t^2)\ge \mu_t^2+\sigma_w^2+\sigma_v^2\\
\E_{{\boldsymbol\theta}_1}(y_t)=  \sum_{k=0}^\infty c_k\mu_{t-k}
\end{align}
Therefore, we have
\begin{align}
\lVert\E_{{\boldsymbol\theta}_1}\tilde{\bf z}_i\rVert\ge\rho\triangleq \left[ \left(m^2+{\sigma_w^2}\right)^2/(2{\sigma_v^4})+m^2/\sigma_v^2\right]^{\frac{1}{2}}
\end{align}
and choosing $b\le 2\rho$ satisfies the second inequality in \eqref{detectability}.

{\color{red}\bf How to choose $h$ and the asymptotic performance.}

Moreover, it is known that, as the detection threshold $h\to \infty$, the following approximation holds \cite{Basseville}:
\begin{align}
\E_{{\boldsymbol\theta}_1}(T)\approx \frac{\ln (\alpha^{-1})}{\E_{\boldsymbol\theta_1} ({\tilde{s}_i})}\leq \frac{\ln (\alpha^{-1})}{b\rho-b^2/2}\;,
\end{align}
where $T$ denotes the detection delay and $\alpha$ is the false alarm probability. Therefore, given the false alarm probability, the mean detection delay is minimized when $\E_{\boldsymbol\theta_1} (\tilde{s}_i^{(\ell)})$ attains its maximum at $b=2\rho$. In practice, there is usually some minimum  disturbance SNR that is of interest, denoted as $\rho_\text{min}$.
%Moreover, recalling that $b$ is, in fact, a small number due to the local assumption we make in Section III.

Therefore, in order to secure the performance in the least favorable case (with smallest SNR), we suggest to select $b=\frac{\rho_\text{min}}{\sqrt{2}}$ such that $\E_{{\boldsymbol\theta}_1}(T)$ is minimized at $\rho_\text{min}$.

In this subsection, we discuss the selection of the free parameters in our proposed detector, i.e., $\{\lD, \uD, b\}$. The interval $[-\lD,\uD]$ controls the communication frequency between the sensors and the central meter. Denote the communication interval as $\tau$. Then we want to select $[-\lD,\uD]$ such that some target value for $\E_{\boldsymbol\theta_0}(\tau)$ is attained, by noting that it is impossible to fix a target value $\E_{\boldsymbol\theta_1}(\tau)$ due to the lack of the prior knowledge of the disturbance.
In \cite{Yasin12}, for the sequential probability ratio test with i.i.d. observations,  the relationship between $[-\lD,\uD]$ and the average transmission interval was derived. However, in our case of correlated observations, closed-form expressions for $\lD$ and $\uD$ are not available in general. Nevertheless, we can adjust $[-\lD, \uD]$  by an off-line stochastic simulation to achieve the target $\E_{\boldsymbol\theta_0}(\tau)$ based on the statistical distribution of the noise. For the experiments in Section V, we set $\lD=\uD$ and adjust the value to meet the target average transmission interval.
}
\subsection{Parameters Tuning}

In this subsection, we discuss the selection of the free parameters in our proposed detector, i.e., $\{\lD, \uD, b\}$. The interval $[-\lD,\uD]$ controls the communication frequency between the distributed meters and the central meter. Denote the communication interval as $\tau$. Then we want to select $[-\lD,\uD]$ such that some target value for $\E_{\boldsymbol\theta_0}(\tau)$ is attained, by noting that it is impossible to fix a target value $\E_{\boldsymbol\theta_1}(\tau)$ due to the lack of the prior knowledge of the disturbance.
%Note that the average transmission intervals before and after the occurrence of disturbances are usually unequal because the distributions  before and after the change point are not symmetric. We suggest to emphasize more on controlling the transmission frequency under the normal condition, while frequent transmission is acceptable after the disturbances occur from a quick detection perspective. In practical sense, we need to keep the sensors less active and energy-saving when  the power quality is normal and switch to active mode adaptively after the occurrence of disturbances.
We can adjust $[-\lD, \uD]$  by an off-line stochastic simulation to achieve the target $\E_{\boldsymbol\theta_0}(\tau)$ based on the statistical distribution of the noise. For the experiments in Section V, we set $\lD=\uD$ and adjust the value to meet the target average transmission interval.

To choose a proper $b$, first we introduce the detectability condition for the general sequential change detection \cite{Basseville}, which states that
%The Kullback-Leibler information between two distributions (parameterized by ${\boldsymbol\theta}_0$ and ${\boldsymbol\theta}_1$) is defined as
%\begin{align}
%  &{K}({\boldsymbol\theta}_0,{\boldsymbol\theta}_1)=\int\log\frac{f_{{\boldsymbol\theta}_0}(y)}{f_{{\boldsymbol\theta}_1}(y)}f_{{\boldsymbol\theta}_0}(y){\rm d}y=-\E_{{\boldsymbol\theta}_0}(s_i),\\
%  &{K}({\boldsymbol\theta}_1,{\boldsymbol\theta}_0)=\int\log\frac{f_{{\boldsymbol\theta}_1}(y)}{f_{{\boldsymbol\theta}_{\boldsymbol\theta}}(y)}f_{{\boldsymbol\theta}_1}(y){\rm d}y=\E_{{\boldsymbol\theta}_1}(s_i).
%\end{align}
the two distributions before and after the occurrence of the disturbance are detectable with finite stopping time if and only if $\E_{{\boldsymbol\theta}_0}(\phi_k)<0<\E_{{\boldsymbol\theta}_1}(\phi_k)$, where $\phi_k$ represents the $k$th sample statistic. This condition implies that the decision statistic $S_1^N=\sum_{k=1}^N\phi_k$ takes different drifting directions between the pre-change and post-change distributions. For the test statistic that cannot be written as the sum of single-sample statistic $\phi_k$, the corresponding quantity is also defined in \cite{Basseville} as $\E_{\boldsymbol\theta}(\phi_k)\triangleq \lim_{N\to \infty}\E_{\boldsymbol\theta}(S_1^N)/N$, where $S_1^N$ is the statistic with $N$ samples.
%$\E_{{\boldsymbol\theta}_0}(\tilde{s}_i^{(\ell)})<0<\E_{{\boldsymbol\theta}_1}(\tilde{s}_i^{(\ell)})$,
It is clear that our statistic in \eqref{eq:decentral_S}-\eqref{eq:individual} corresponds to the latter definition. Thus at the $\ell$th meter we have
\ignore{\begin{align}
\E_{\boldsymbol\theta}(\tilde{\phi}_k^{(\ell)})&\triangleq \lim_{(k-j+1)\to \infty}\frac{1}{k-j+1}\E_{\boldsymbol\theta}(\tilde{S}_j^{k, (\ell)})\nonumber\\
&=\lim_{(k-j+1)\to\infty}\frac{b}{k-j+1}\E_{\boldsymbol\theta}({U_j^k}^{(\ell)})-\frac{b^2}{2},\quad \nonumber\\ &\qquad\qquad \qquad  \quad {\boldsymbol\theta}= {\boldsymbol\theta}_0, {\boldsymbol\theta}_1\;, \ell=1, 2, \ldots, L.
\end{align}
Substituting ${U_j^k}^{(\ell)}$ with \eqref{eq:individual}, we obtain
\begin{align}
  &\lim_{(k-j+1)\to \infty}\frac{b}{k-j+1}\E_{\boldsymbol\theta}({U_j^k}^{(\ell)})\nonumber\\&=b\;\E_{\boldsymbol\theta}\left(\sqrt{\sum_{m=1}^p\left(\frac{\sum_{i=j}^ky_i^{(\ell)}y^{(\ell)}_{i-m}}{(k-j+1)\sigma_\nu^2}\right)^2+\frac{1}{2}\left(\frac{\sum_{i=j}^k{(y_i^{(\ell)})}^2/\sigma_\nu^2}{k-j+1}-1\right)^2+\left(\frac{\sum_{i=j}^ky_i^{(\ell)}}{k-j+1}/\sigma_\nu\right)^2}\right)\nonumber\\&\to
  {b}\rho\;,\label{eq:E_H}
\end{align}}
\begin{align}
\E_{\boldsymbol\theta}(\tilde{\phi}_k^{(\ell)})&\triangleq \lim_{N\to \infty}\frac{1}{N}\E_{\boldsymbol\theta}(\tilde{S}_1^{N, (\ell)})\nonumber\\
&=\lim_{N\to\infty}\frac{b}{N}\E_{\boldsymbol\theta}({U_1^N}^{(\ell)})-\frac{b^2}{2},\quad \nonumber\\ &\qquad\qquad \qquad  \quad {\boldsymbol\theta}= {\boldsymbol\theta}_0, {\boldsymbol\theta}_1\;, \ell=1, 2, \ldots, L.
\end{align}
Substituting ${U_j^k}^{(\ell)}$ with \eqref{eq:individual}, we obtain
\ignore{\begin{align}
  &\lim_{(k-j+1)\to \infty}\frac{b}{k-j+1}\E_{\boldsymbol\theta}({U_j^k}^{(\ell)})\nonumber\\&=b\;\E_{\boldsymbol\theta}\left(\sqrt{\sum_{m=1}^p\left(\frac{\sum_{i=j}^ky_i^{(\ell)}y^{(\ell)}_{i-m}}{(k-j+1)\sigma_\nu^2}\right)^2+\frac{1}{2}\left(\frac{\sum_{i=j}^k{(y_i^{(\ell)})}^2/\sigma_\nu^2}{k-j+1}-1\right)^2+\left(\frac{\sum_{i=j}^ky_i^{(\ell)}}{k-j+1}/\sigma_\nu\right)^2}\right)\nonumber\\&\to
  {b}\rho\;,\label{eq:E_H}
\end{align}}
\begin{align}
  &\lim_{N\to \infty}\frac{b}{N}\E_{\boldsymbol\theta}({U_1^N}^{(\ell)})\nonumber\\&=b\;\E_{\boldsymbol\theta}\Bigg(\lim_{N\to \infty}\Bigg\{\sum_{m=1}^p\left(\frac{\sum_{i=1}^Ny_i^{(\ell)}y^{(\ell)}_{i-m}}{N\sigma_\nu^2}\right)^2+\frac{1}{2}\left(\frac{\sum_{i=1}^N{(y_i^{(\ell)})}^2}{N\sigma_\nu^2}-1\right)^2+\left(\frac{\sum_{i=1}^Ny_i^{(\ell)}}{N\sigma_\nu}\right)^2\Bigg\}^{1/2}\Bigg)\nonumber\\&=
  {b}\rho^{(\ell)}\;,\label{eq:E_H}
\end{align}
where
\begin{align}
&\rho^{(\ell)}\triangleq \left\{\!\sum_{m=1}^p\left(\frac{R^{\ell}(m)\!+\!(\tilde{\mu}^{(\ell)})^2}{\sigma_\nu^2}\right)^2\!+\left(\frac{R^{\ell}(0)\!+\!(\tilde{\mu}^{(\ell)})^2-\sigma_\nu^2}{\sqrt{2}\sigma_\nu^2}\right)^2\!\!+\!(\tilde{\mu}^{(\ell)})^2/\sigma_\nu^2\right\}^{1/2},
\end{align}
$R^\ell(m),\; m=0, 1, \ldots, p$ and $\tilde{\mu}^{(\ell)}$ are the autocovariance and the mean shift of the observations at meter $\ell$ respectively. 
%Note that $\rho^{(\ell)}$ is determined by the disturbance signal at sensor $\ell$.

We choose $b$ such that the detectability condition holds, i.e., $\E_{{\boldsymbol\theta}_0}(\tilde{\phi}^{(\ell)}_k)<0<\E_{{\boldsymbol\theta}_1}(\tilde{\phi}^{(\ell)}_k),\; \ell=1, 2, \ldots, L$. Specifically, before the disturbance occurs, only the white Gaussian noise exists, hence $\tilde{\mu}^{(\ell)}=0, R^\ell(0)=\sigma_\nu^2, \; R^\ell(1)=\cdots=R^\ell(p)=0$, thus we have
\begin{align}\label{local-LLR-0}
\E_{\boldsymbol\theta_0}(\tilde{\phi}_k^{(\ell)})=-\frac{b^2}{2}<0, \quad \ell=1, 2, \ldots, L,
\end{align}
for any positive value of $b$.
After the disturbance occurs, we have
\begin{align}\label{eq:E_S}
\E_{\boldsymbol\theta_1} (\tilde{\phi}_k^{(\ell)})&\!=b\rho^{(\ell)}-\frac{b^2}{2}>0, \quad  \ell=1, 2, \ldots, L,
\end{align}
for $0<b<2\min \left\{\rho^{(\ell)}, \ell=1, \ldots, L\right\}$.
Furthermore, it is known that, as the detection threshold $h\to \infty$, the following approximation holds \cite{Basseville}:
\begin{align}
\E_{{\boldsymbol\theta}_1}(T)\approx \frac{\ln \gamma}{\sum_{\ell=1}^L\E_{\boldsymbol\theta_1} (\tilde{\phi}_k^{(\ell)})}\leq \frac{\ln \gamma}{b\sum_{\ell=1}^L\rho^{(\ell)}-b^2L/2}\;,
\end{align}
where $T$ denotes the detection delay and $\gamma$ is the false alarm period. Therefore, given the false alarm period, the mean detection delay is minimized when $\sum_{\ell=1}^L\E_{\boldsymbol\theta_1} (\tilde{\phi}_k^{(\ell)})$ attains its maximum at $b=\frac{1}{L}\sum_{\ell=1}^L\rho^{(\ell)}$. In practice, the practitioners can decide the minimum $\rho_\text{min}^{(\ell)}$ at each meter and select $b=\frac{1}{L}\sum_{\ell=1}^L\rho_\text{min}^{(\ell)}$ such that the detection performance for worst-case disturbance is optimized.
%Therefore, in order to secure the performance in the least favorable case (with smallest SNR), we suggest to select $b=\frac{\rho_\text{min}}{\sqrt{2}}$ such that $\E_{{\boldsymbol\theta}_1}(T)$ is minimized at $\rho_\text{min}$.

\section{Simulation Results}
In this section, we first apply the proposed detector on some typical power quality disturbances to demonstrate that it promptly detects the occurrence of these disturbances. In specific, we compare the proposed detector with the widely used RMS method, the STFT method and the weighted CUSUM method in \cite{Xingze10}. Then we examine the performance of the proposed cooperative detection scheme as well as its decentralized implementation based on the level-triggered sampling.

\begin{figure}
\centering
\includegraphics[width=0.99\textwidth]{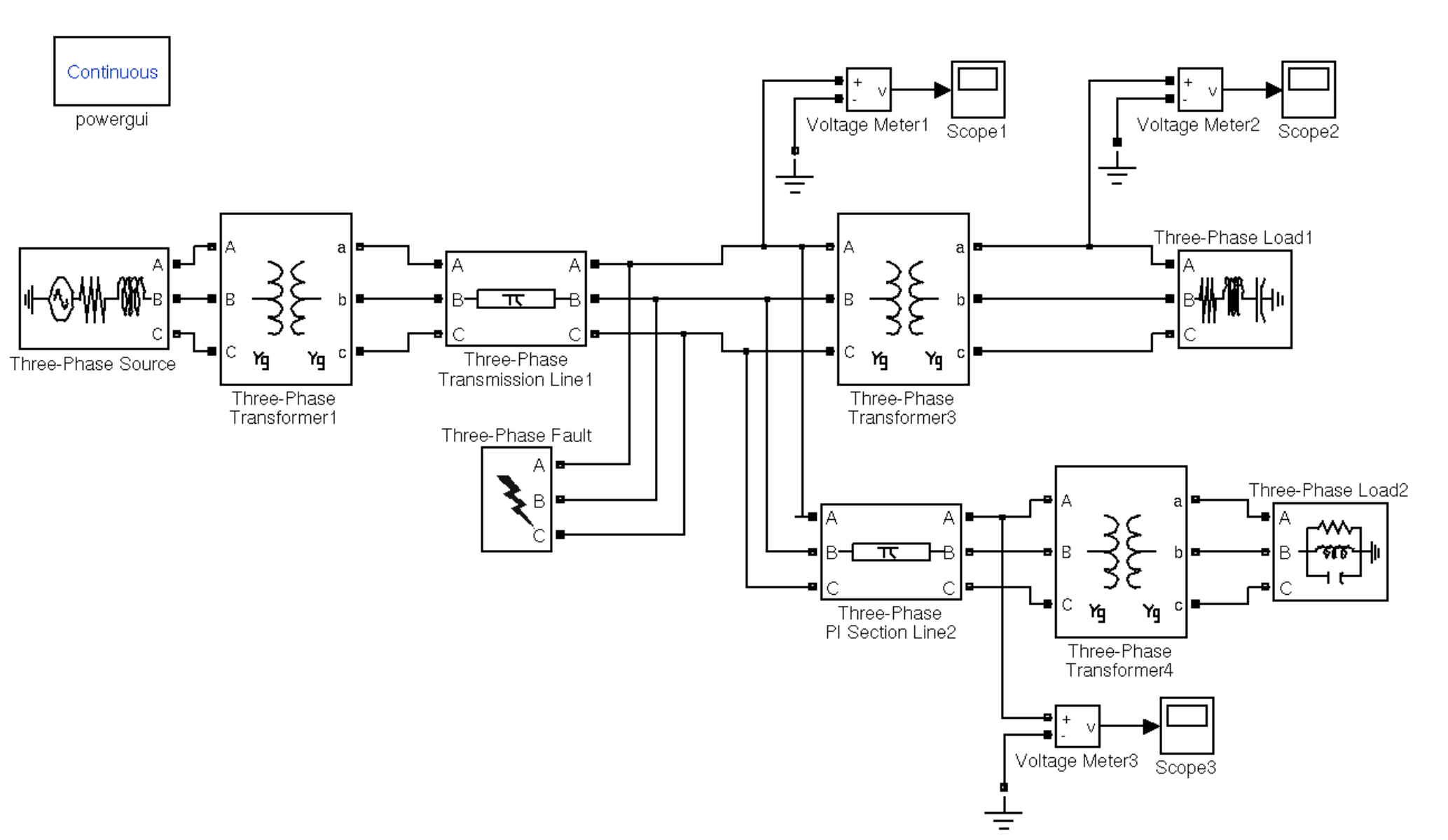}
\caption{The simulation system for fault-induced power sag disturbance.}\label{fig:fault}
\end{figure}

\begin{figure}
\centering
\includegraphics[width=0.99\textwidth]{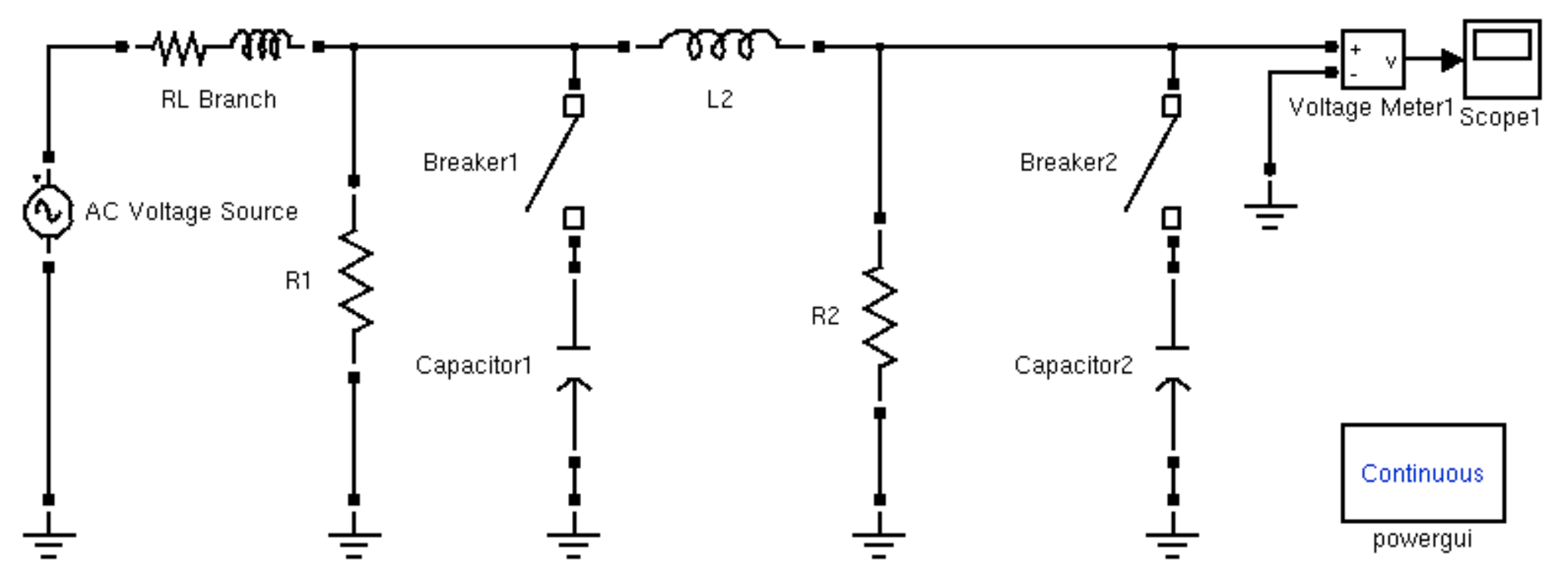}
\caption{The simulation system for capacitor-switching-induced transient power disturbance.}\label{fig:transient}
\end{figure}

In our experiment, the disturbance signals are obtained by constructing  simulation systems using the popular Matlab toolbox SimPowerSystems \cite{SimPower}. We mainly consider the disturbance of voltage sag induced by a distribution line fault (simulated by constructing the network in Fig. \ref{fig:fault} according to \cite{Tan13}) and the transient disturbance induced by the capacitor bank switching (simulated by constructing the network in Fig. \ref{fig:transient} according to \cite{CapacitorBank}). In particular, Fig. \ref{fig:fault} corresponds to a simplified distribution network where three-phase power supply is transmitted and distributed to ``load 1'' and ``load 2''. Three meters are deployed to monitor this distribution network. In Fig. \ref{fig:transient}, the ``capacitor 1'' and ``capacitor 2'' constitute the capacitor bank which can be switched on and off to adjust the power factor. Throughout the experiment, The nominal voltage is a sinusoidal waveform with $f_0=60$Hz and unit magnitude. The sampling rates at all meters are set as the standard $64$ samples per cycle, i.e., sampling frequency $f_s={60\times 64}$ Hz. The GLLR detector is applied with a first-order AR model (i.e., $p=1$) and the parameter is set as $b=0.5$.

\subsection{Comparison with Existing Methods (Single Meter)}

\begin{figure}
\centering
\includegraphics[width=0.99\textwidth]{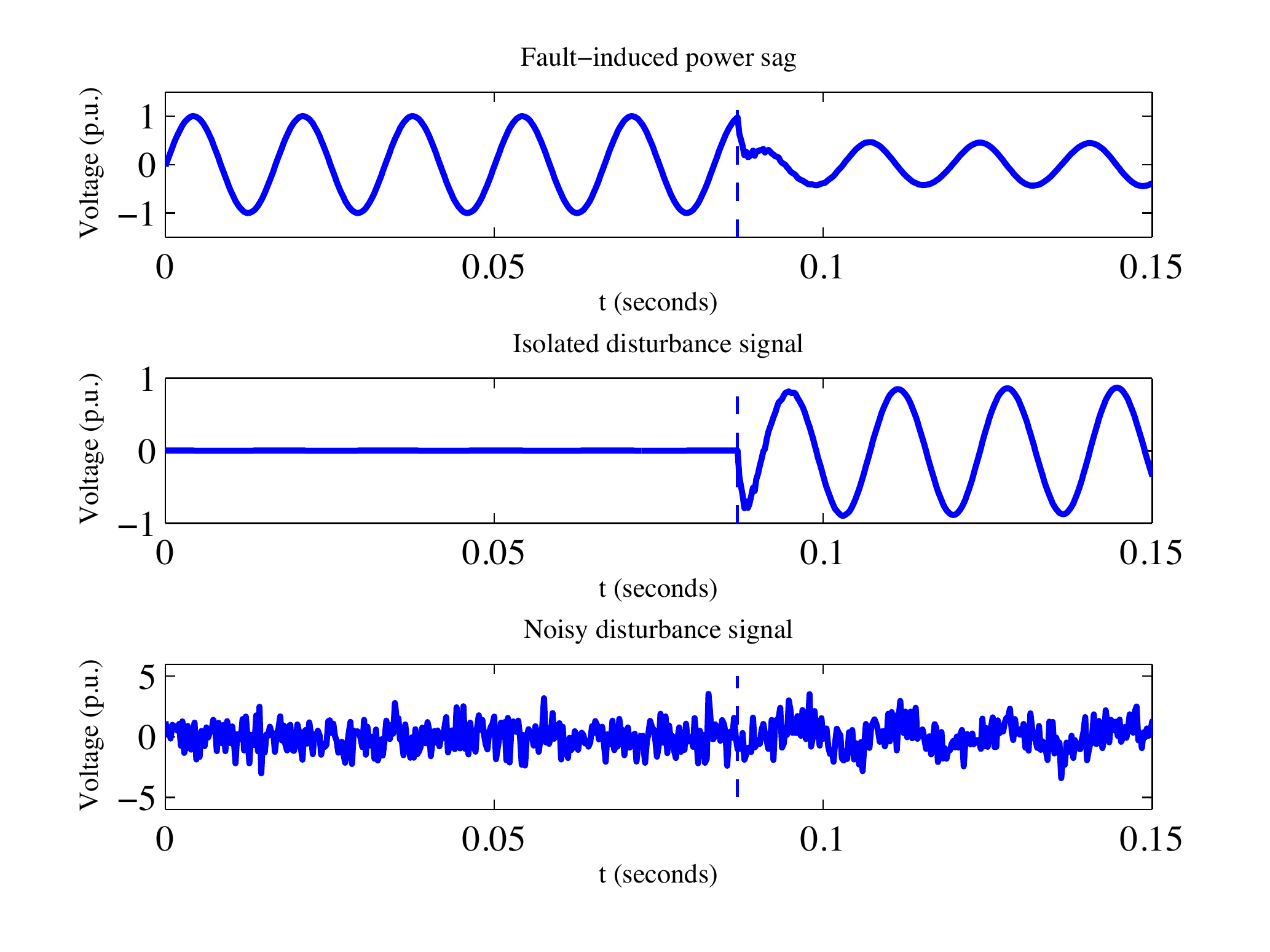}
\caption{Original voltage waveform with power sag, isolated disturbance signal and the corresponding noisy measurements.}\label{fig:fault_disturbance}
\end{figure}

\begin{figure}
\centering
\includegraphics[width=0.99\textwidth]{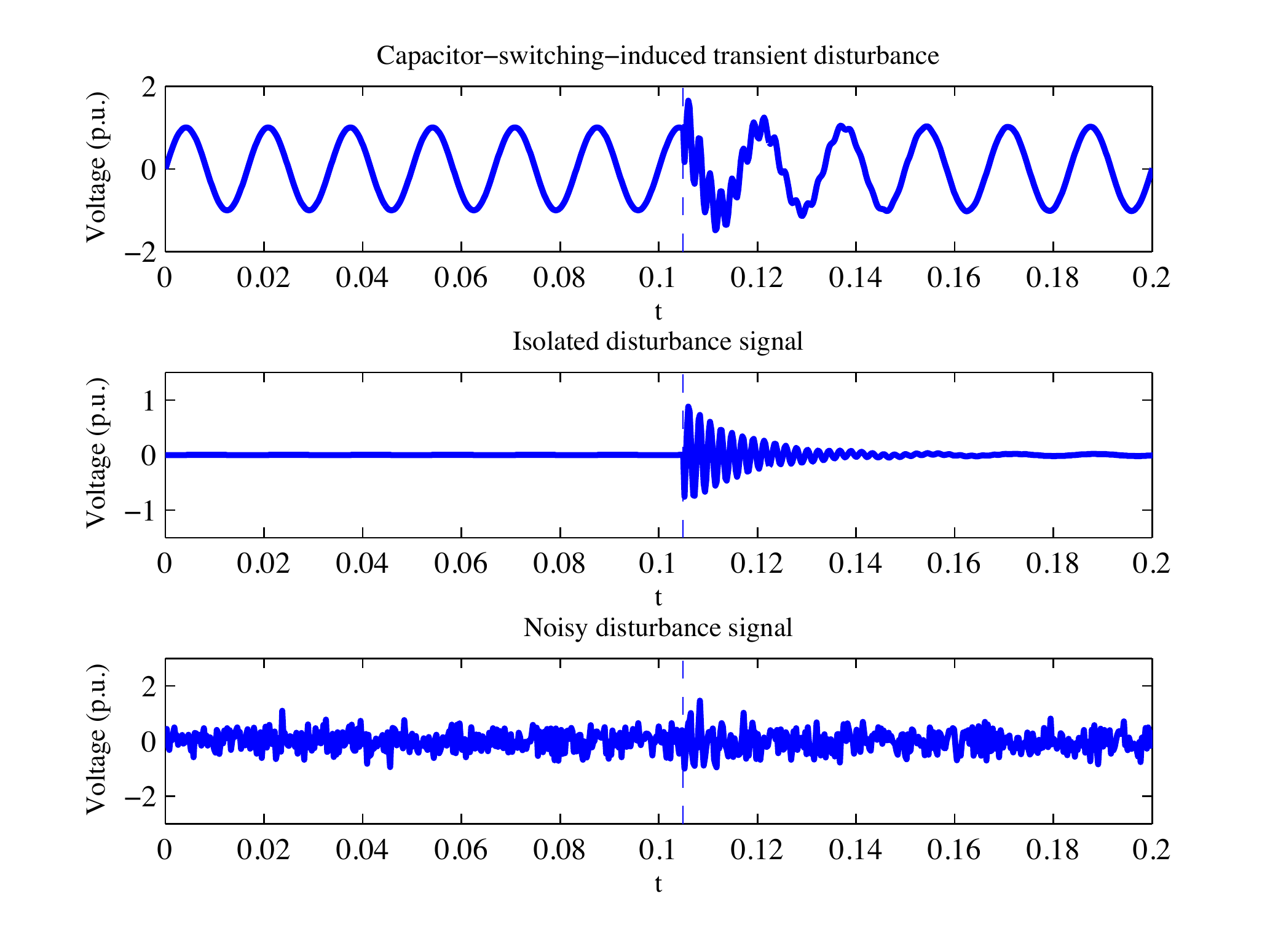}
\caption{Original voltage waveform with transient distortion, isolated disturbance signal and the corresponding noisy measurements.}\label{fig:transient_disturbance}
\end{figure}

By focusing on the single-meter detection, we compare the proposed GLLR detector with the widely adopted methods, namely the RMS method, the STFT method and the weighted CUSUM test. Fig. \ref{fig:fault_disturbance} illustrates the power sag disturbance incurred by the ``Phase A line'' fault at $t=0.0869s$ (the occurrence is marked with dashed blue line). Fig. \ref{fig:transient_disturbance} illustrates the power transient distortions incurred by the closing ``capacitor 1'' in Fig. \ref{fig:transient} at $t=0.105s$. They correspond to the voltage waveform at ``Meter 1'' in Fig. \ref{fig:fault} and Fig. \ref{fig:transient} respectively. Both the RMS method and the STFT method are implemented with a one-cycle window that slides point by point, achieving the best possible time resolution. Thus the STFT method performs the $64$-point FFT within each window. One can chose a larger window for higher resolution in the frequency domain at the price of lower resolution in the time domain, which is less desirable when quick detection is considered. The weighted CUSUM is implemented using the Gaussian prior as proposed in \cite{Xingze10}.


\begin{figure}
\centering
\includegraphics[width=0.49\textwidth]{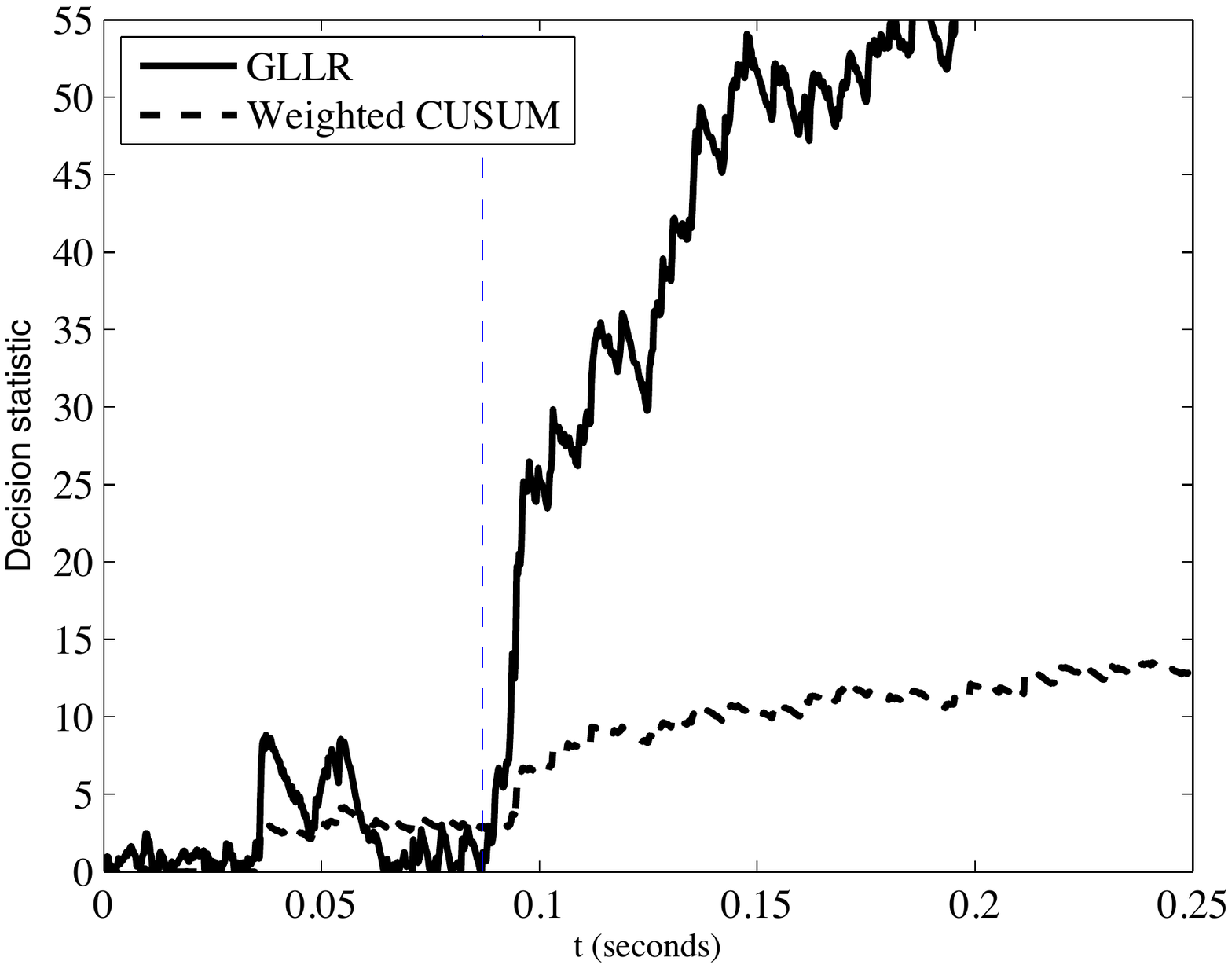}
\includegraphics[width=0.49\textwidth]{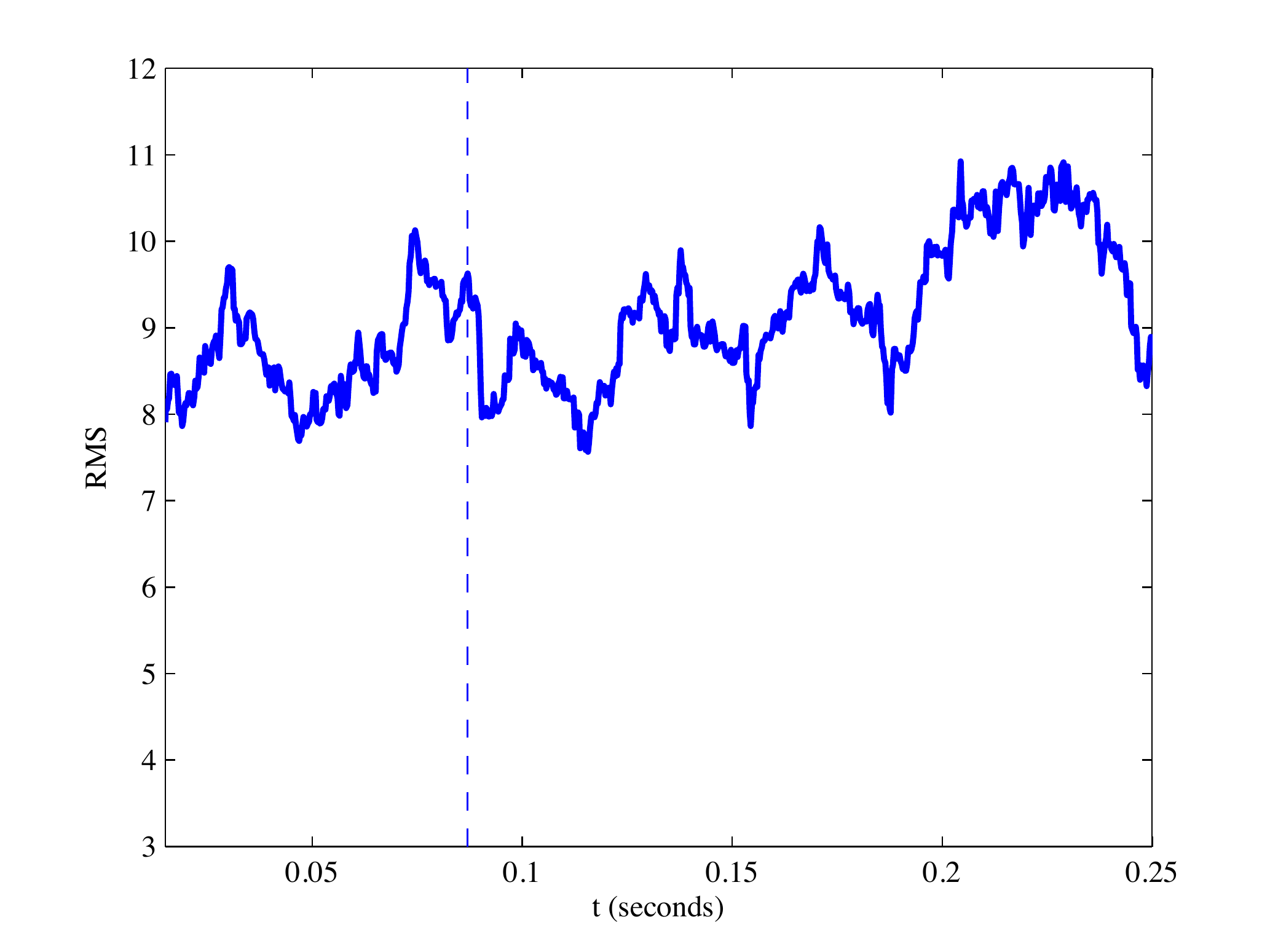}
\includegraphics[width=0.49\textwidth]{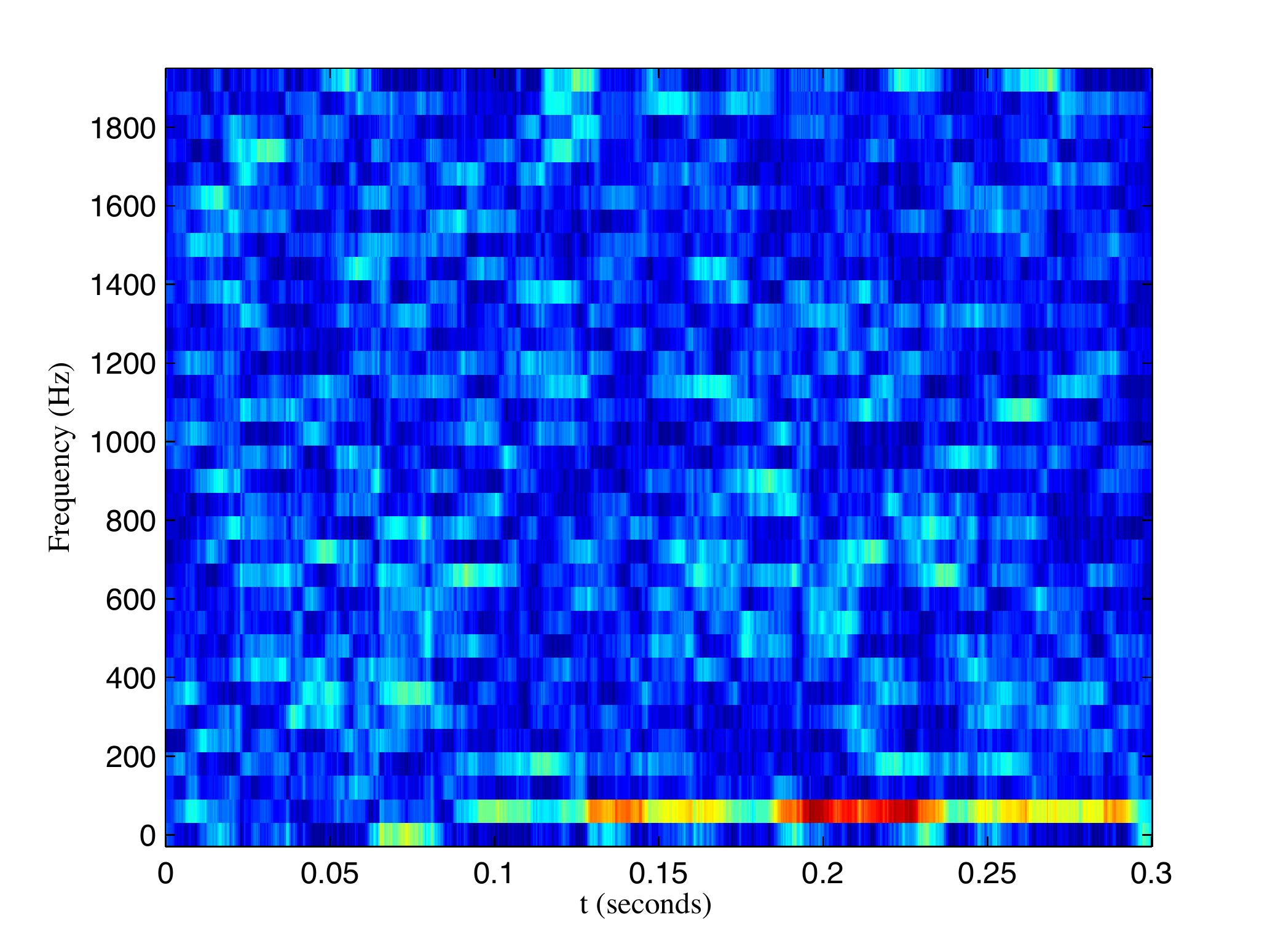}
\caption{Detection of power sag disturbance using the GLLR, weighted CUSUM, RMS and STFT methods.}\label{fig:ds_fault}
\end{figure}

\begin{figure}
\centering
\includegraphics[width=0.49\textwidth]{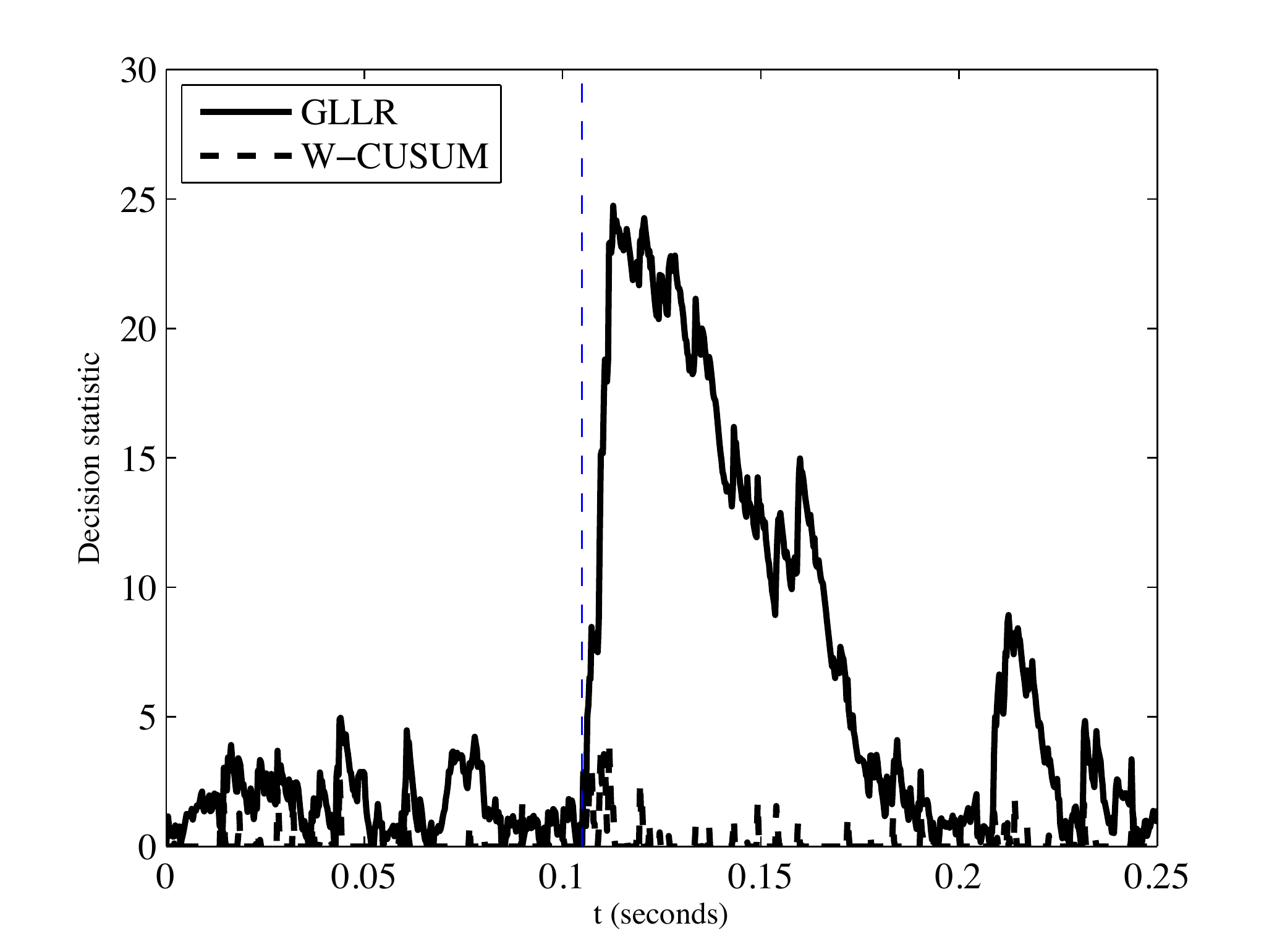}
\includegraphics[width=0.49\textwidth]{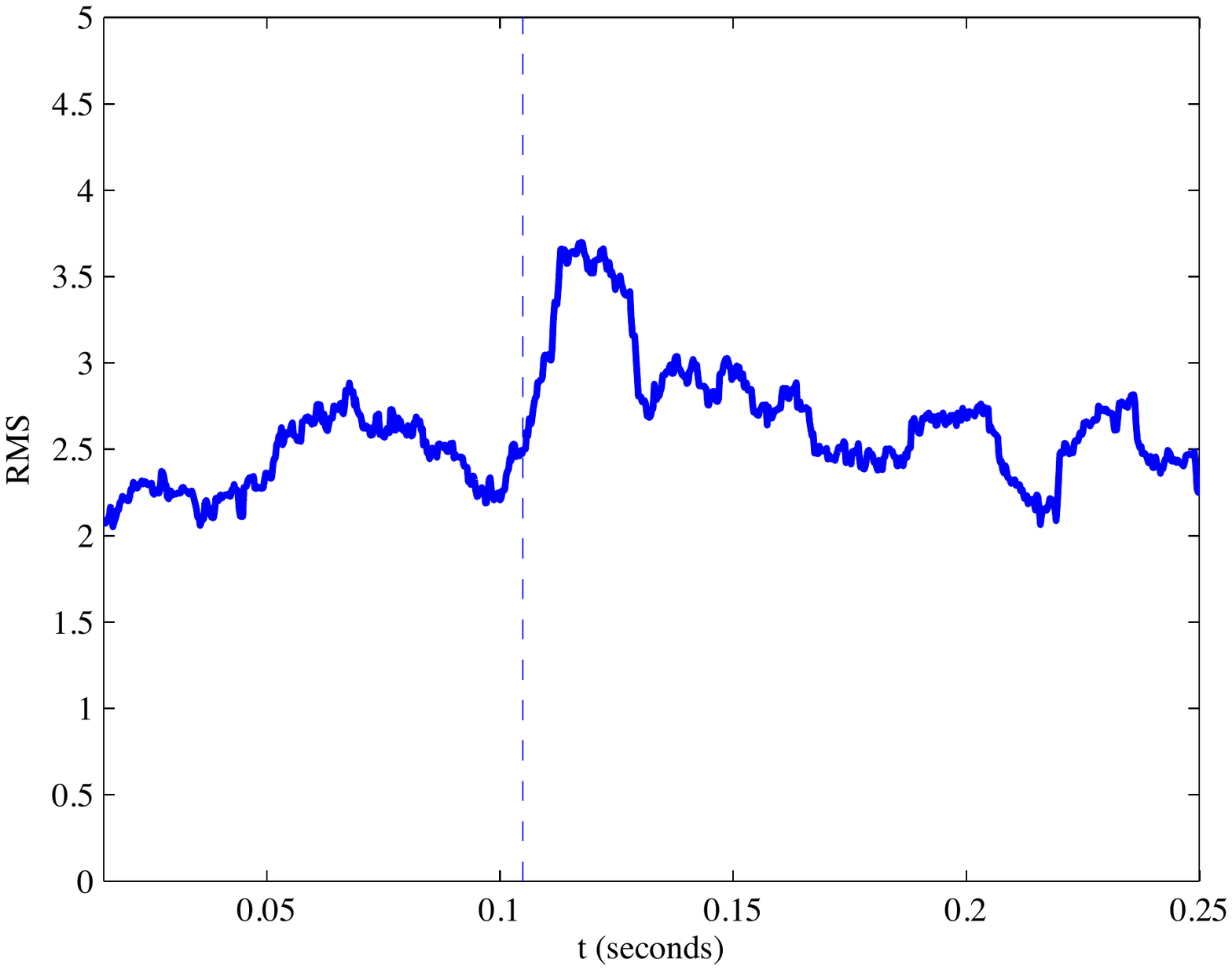}
\includegraphics[width=0.49\textwidth]{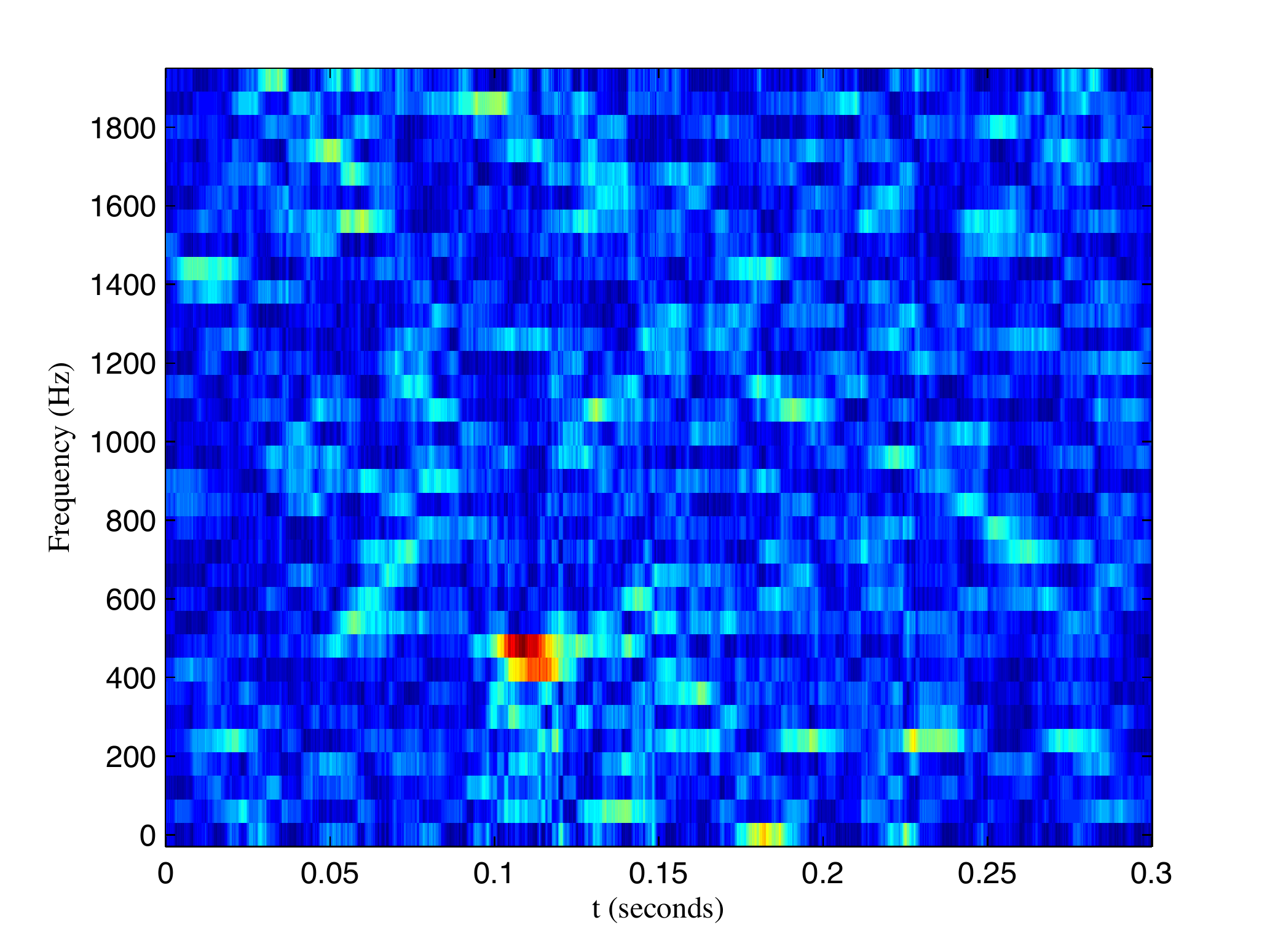}
\caption{Detection of transient disturbance using the GLLR, weighted CUSUM, RMS and STFT methods. }\label{fig:ds_transient}
\end{figure}

Figs. \ref{fig:ds_fault}-\ref{fig:ds_transient} show the decision statistics of the proposed GLLR detector, weighted CUSUM (W-CUSUM), RMS detector and STFT detector. For both types of disturbances, the decision statistic of GLLR exhibits abrupt changes on the occurrence of the disturbance. In contrast, the decision statistic of W-CUSUM slowly increases after the occurrence of power sag and fails to detect the transient disturbance. Moreover, in both cases, the STFT detector vaguely shows the presence of new frequency component upon the occurrence of disturbance, and the RMS detector does not indicate the occurrence clearly.

\begin{figure}
\centering
\includegraphics[width=0.95\textwidth]{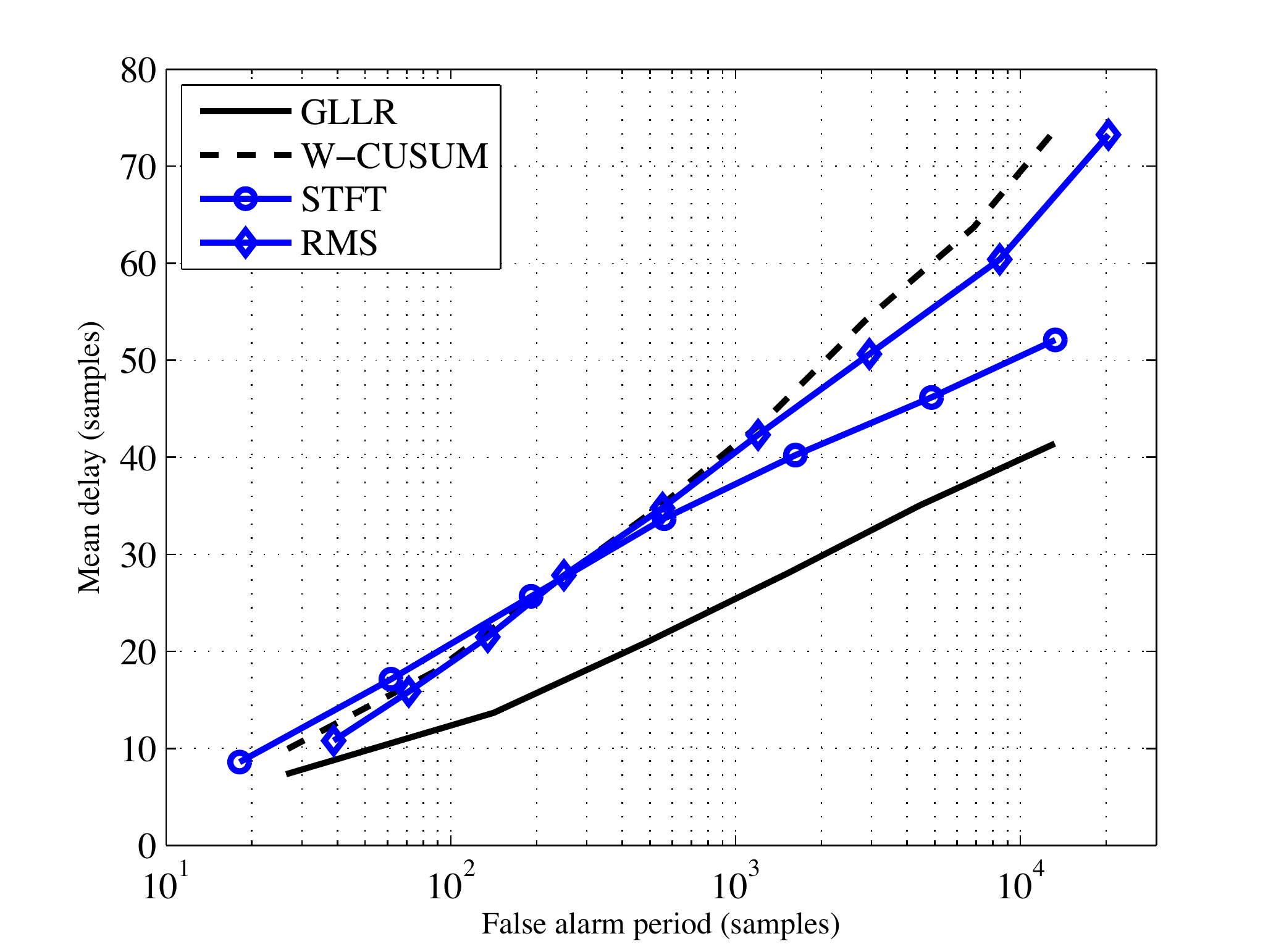}
\caption{The detection delay versus the false alarm period for the GLLR detector, weighted CUSUM detector, STFT method and RMS method.}\label{fig:Flp_Delay_S}
\end{figure}

To perform a rigorous comparison, the mean detection delay versus the false alarm period is further examined based on the power sag disturbance shown in Fig. \ref{fig:Flp_Delay_S}. It is seen that the GLLR outperforms the other methods by yielding much shorter mean delay with the same false alarm period. That is, the GLLR detector reacts much quicker to the occurrence of disturbance signal.

\subsection{Cooperative Detector (Centralized and Decentralized)}

\begin{figure}
\centering
\includegraphics[width=0.99\textwidth]{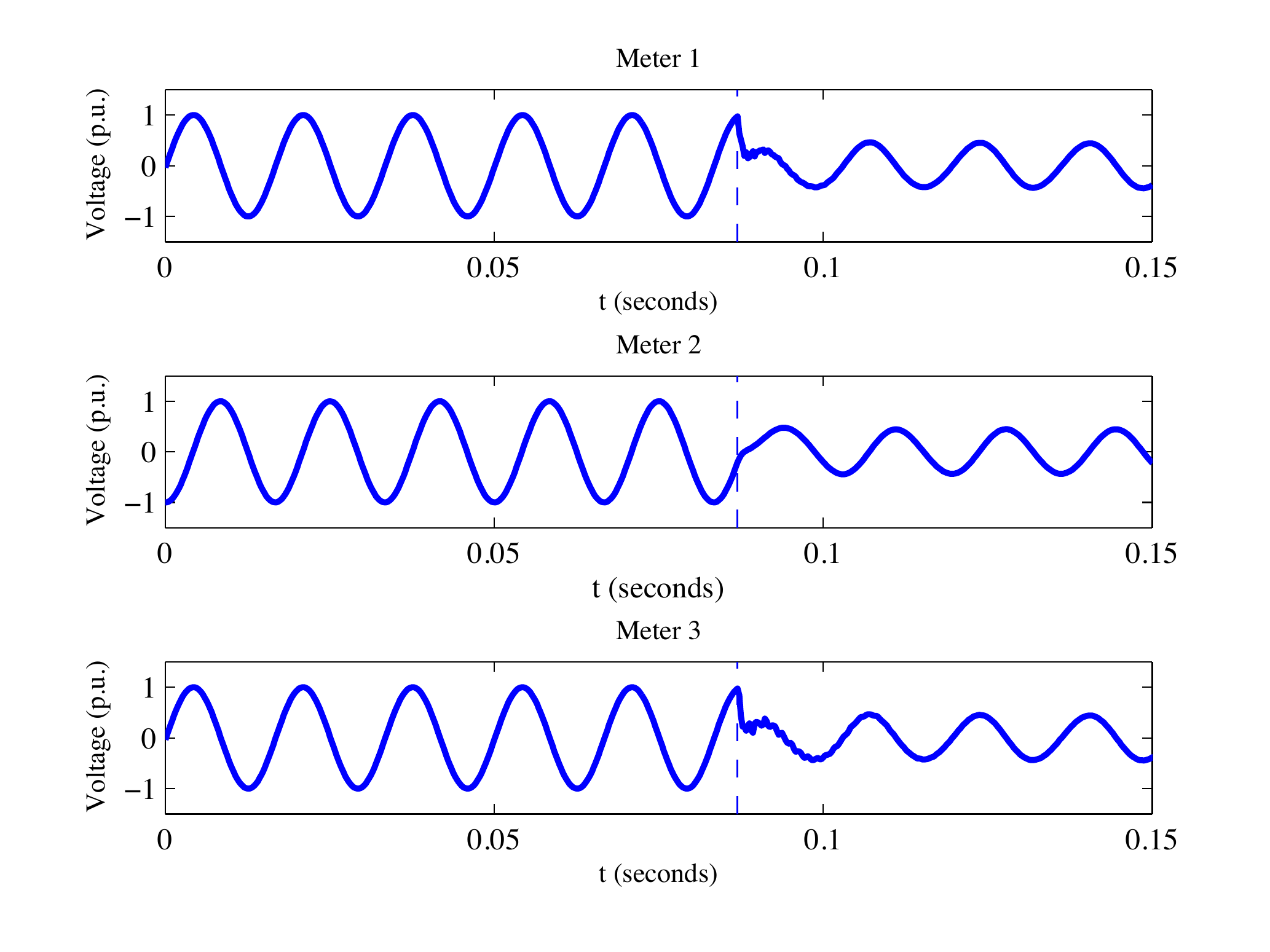}
\caption{Original voltage waveform with fault-induced power sag disturbance at Meters 1-3.}\label{fig:fault_M}
\end{figure}

\begin{figure}
\centering
\includegraphics[width=0.95\textwidth]{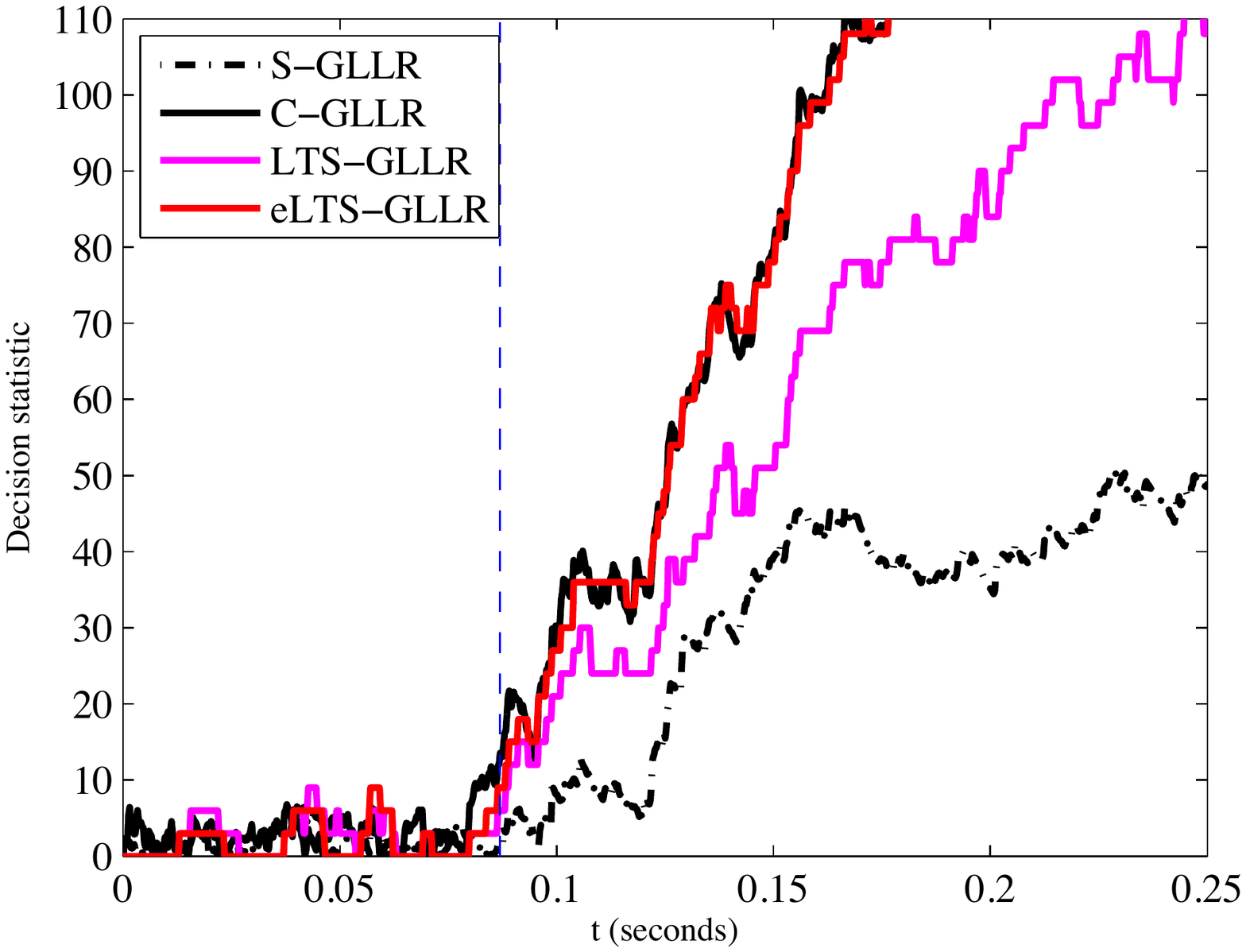}
%\caption{Noise level $\sigma_n^2=1$. $\Delta=3$. $b=0.5$. Three sensors.}
\caption{Decision statistics of the single-meter detector and the cooperative detectors based on level-triggered sampling. }
\label{fig:ds_M}
\end{figure}

We next incorporate more meters in the network and examine the performance of  cooperative detection. Focusing on the power sag event, the disturbance signals observed at Meters 1-3 are illustrated in Fig. \ref{fig:fault_M}. We see that the disturbance signals induced by the same event occur at the same time to multiple buses but vary from each other in terms of the waveform.
%The decentralized GLLR detector based on level-triggered sampling (LTS-GLLR) and its enhanced version (eLTS-GLLR) are implemented with $\uD=\lD=3$.
In Fig. \ref{fig:ds_M}, the decision statistics of the single-meter detector (i.e., S-GLLR), the centralized cooperative detector  (i.e., C-GLLR), the decentralized detector based on level-triggered sampling (LTS-GLLR) and the enhanced LTS-GLLR (eLTS-GLLR) are plotted. First, the cooperative detector exhibits steeper increase of the decision statistic compared to the single-meter detector, implying a more prompt reaction to disturbance signals. In the mean time, the global decision statistics of LTS-based decentralized detectors are updated with a much lower frequency than the centralized one. In particular, the original LTS-GLLR detector clearly diverges from the centralized one due to overshoot accumulation over time, while the enhanced decentralized detector matches closely with the centralized detector.
Next we examine the cooperative detectors in terms of detection delay versus the false alarm period. The local thresholds for the level-triggered sampling is chosen as $[-\lD, \uD]=[-1.6, 1.6]$, under which, at each distributed meter, we have $\E_{\boldsymbol\theta_0}(\tau)=14$ samples under normal condition, and $\E_{\boldsymbol\theta_1}(\tau)=4$ samples after the occurrence of disturbance. Compared with the single-meter case, it is seen that cooperative detection with three meters substantially improves the performance in terms of achieving a shorter detection delay. Notably, the proposed LTS-based decentralized detector only exhibits a minor increase of detection delay compared to the centralized detector. As expected, the improvement of eLTS-GLLR over the original LTS-GLLR becomes more significant as the detection delay grows and overshoot errors accumulate.

\begin{figure}
\centering
\includegraphics[width=0.95\textwidth]{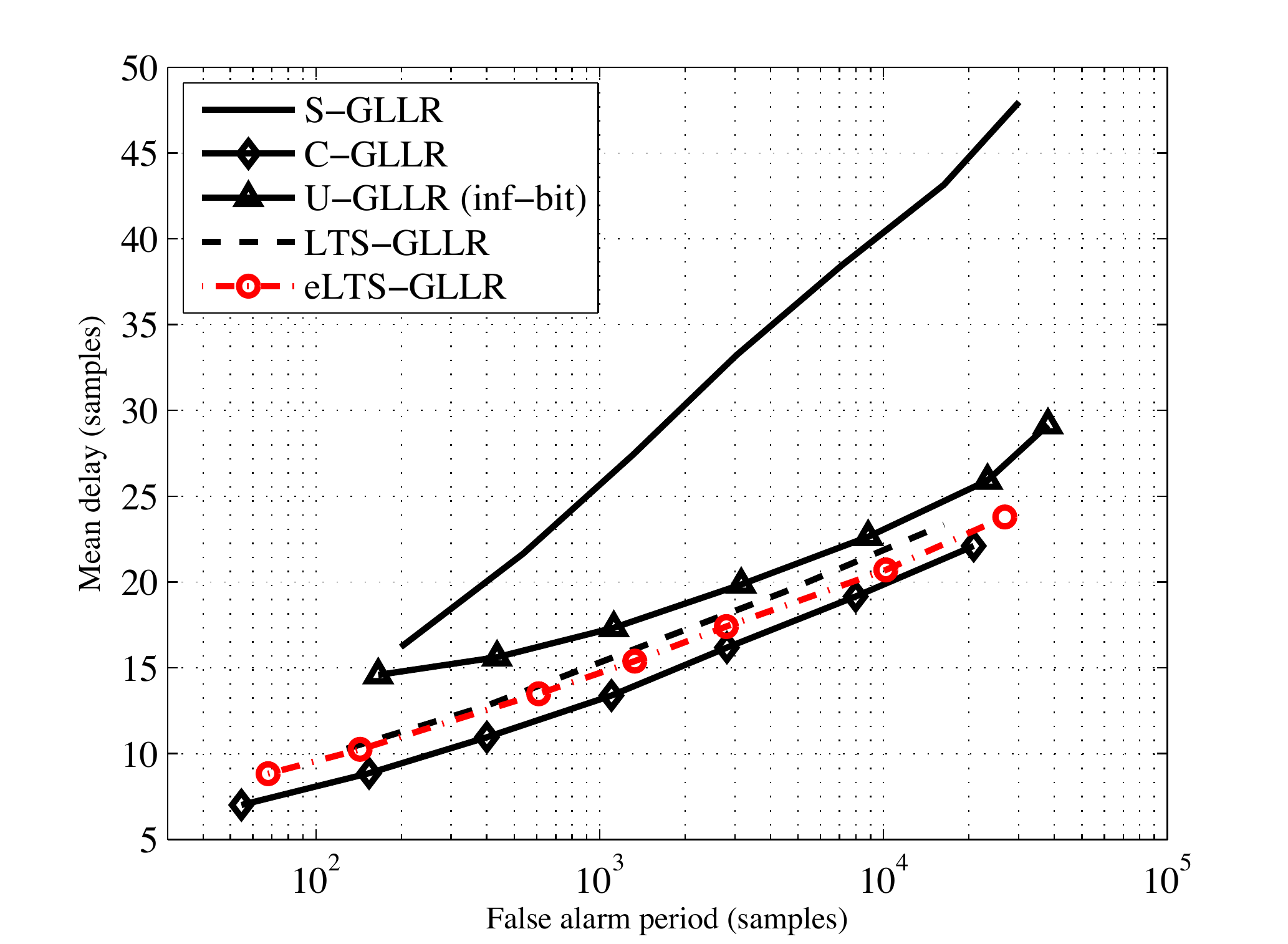}
\caption{The detection delay versus the false alarm period for the single-meter detection and the cooperative detection (centralized and decentralized detectors).}
%\caption{The detection delay versus the false alarm period for the single-meter detection, cooperative detection (centralized and decentralized detectors). Noise level $\sigma_n^2=0.5$. $b=0.5$. $\Delta=1.6$. $T0=14, T1=4$}
\label{fig:Flp_Delay_M}
\end{figure}

In Fig. \ref{fig:Flp_Delay_M}, we also demonstrate the power of the level-triggered sampling by comparing the proposed decentralized detector with a simple decentralized detector, where each local meter computes its local statistic and  transmits it to the central meter every $\tau>1$ sampling instants (also termed as uniform decentralized detector, which we refer to as U-GLLR in the experiment). Note that when $\tau=1$, this scheme becomes the centralized detector.  Here we set $\tau=14$ for the simple decentralized detector to match that of the eLTS-GLLR under normal condition.
%The average communication interval at all three meters for the enhanced LTS-GLLR detector is controlled at $\E_{\boldsymbol\theta_0}(\tau)=14$ and $\E_{\boldsymbol\theta_1}(\tau)=4$.
That is, the simple decentralized detector transmits equally frequently as eLTS-GLLR under the normal condition. However, due to the lack of adaptiveness, the time resolution of U-GLLR is limited by $\tau$ even in the presence of disturbance signals. Moreover, we assume that in the simple decentralized detector, each local meter transmits the exact value of its local statistic which corresponds to infinite number of bits for each transmission; whereas in the decentralized detectors based on level-triggered sampling, only one bit is sent at each transmission. Remarkably, it is seen in Fig. \ref{fig:Flp_Delay_M} that even with only one-bit transmission, level-triggered sampling still outperforms the traditional uniform-in-time sampling that transmits infinite number of bits.

\begin{figure}
\centering
\subfigure[$\sigma_\nu^2=0.5$]{
\includegraphics[width=0.8\textwidth]{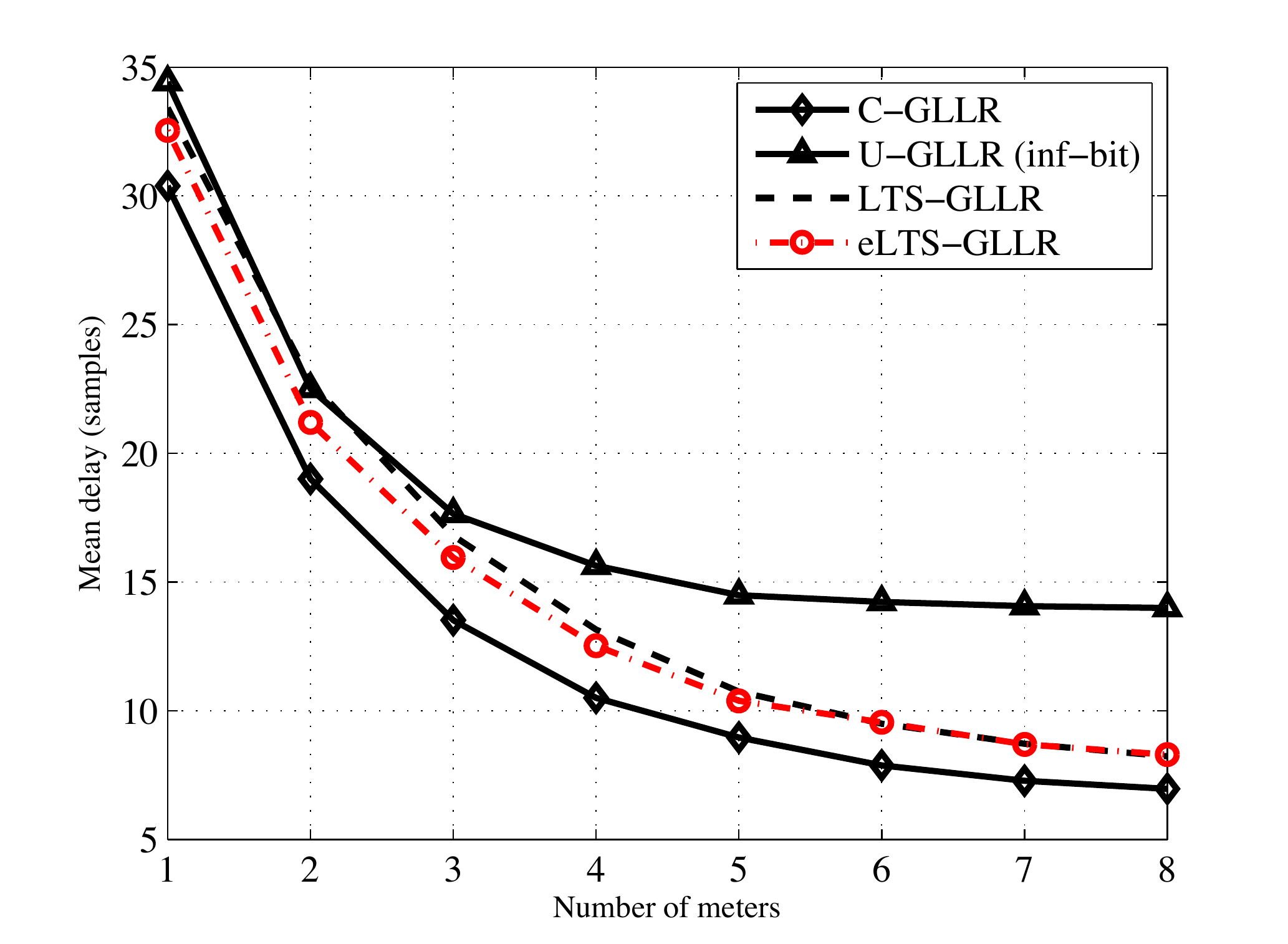}\label{fig:diffK_a}}
\subfigure[$\sigma_\nu^2=1$]{
\includegraphics[width=0.8\textwidth]{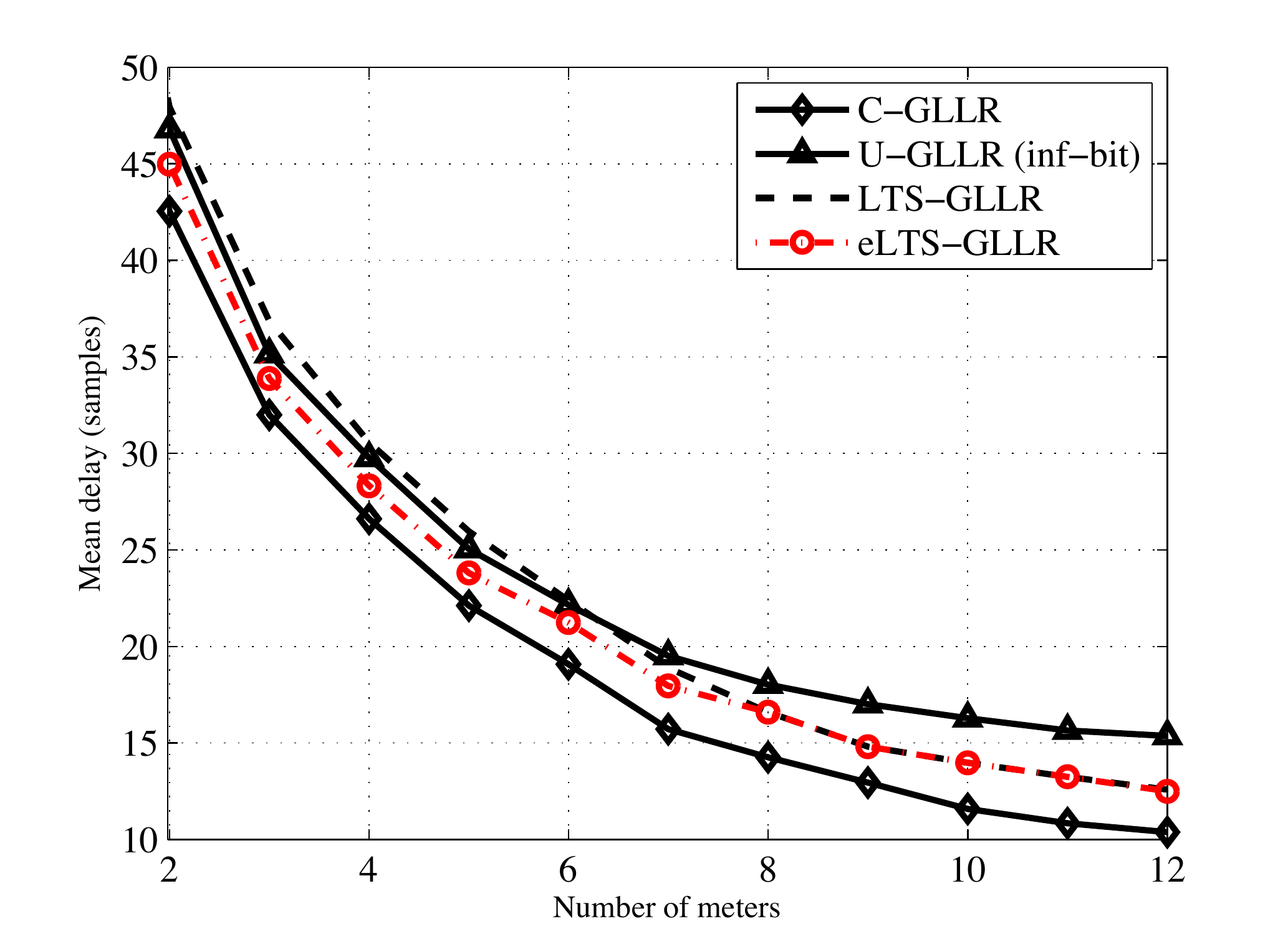}\label{fig:diffK_b}}
\caption{The mean detection delay versus increasing number of meters given $\gamma=2000$.}
%\caption{The mean detection delay versus increasing number of meters given $\gamma=2000$. $\sigma_n^2=1$. $\Delta=1.6$.}
\end{figure}

Finally, Fig. \ref{fig:diffK_a}-(b) depict the performances of the centralized and decentralized detectors as the number of meters grows. Again, we consider the power sag event. The communication rate of the simple decentralized detector is fixed at $\tau=14$, and the average communication rate of proposed decentralized detector is controlled to be $\E_{\boldsymbol\theta_0}(\tau)=14$ and $\E_{\boldsymbol\theta_1}(\tau)=4$. The false alarm period is set as $\gamma = 2\times 10^{3}$. It is seen that the detection performance is significantly improved (i.e., smaller mean delay) as the number of meters grows for all methods, implying the benefit of cooperative detection. As the noise level increases, more distributed meters are required to achieve the same detection performance.
%Note that the simple decentralized detector based on uniform sampling that transmits infinite number of bits consistently exhibits a larger delay than the  proposed decentralized detector based on level-triggered sampling and one-bit transmission.
Note that, due to the lack of adaptiveness, the detection delay of U-GLLR is saturated at mean delay of $14$ samples regardless of increasing number of meters. In contrast, eLTS-GLLR consistently outperforms other decentralized detectors.

\section{Conclusions}
We have developed a cooperative sequential change detection framework for online power quality monitoring. Specifically, local meters observe the voltage signal  independently and communicate wirelessly with a central meter to detect the disturbance. The goal is to achieve the quickest detection under a certain false alarm constraint. First, based on the AR modeling of the disturbance and the sequential change detection framework, we have proposed a sequential GLLR test that does not require the knowledge of the model parameters. Unlike the conventional RMS or STFT method, the proposed technique exploits the statistical distributions of the observed waveform before and after the occurrence of disturbance, thus provides superior performance, especially in the noisy environment. We have also developed the decentralized version of the GLLR detector, which is specifically tailored toward the low-bandwidth requirement imposed by the wireless transmissions between the distributed meters and the central meter. This is achieved by a novel level-triggered sampling scheme that features single-bit information transmission. Finally we have provided extensive simulation results to demonstrate the superior performance of the proposed centralized and decentralized cooperative detectors over the existing methods.
\bibliographystyle{IEEEtran}
\bibliography{IEEEabrv,references}
\end{document}